\documentclass[journal,onecolumn]{IEEEtran}
\usepackage{lineno,hyperref}
\usepackage{fix-cm}
\usepackage{mathtools}
\usepackage{amsmath}
\usepackage{graphicx}
\usepackage{latexsym,enumerate,amssymb,amsbsy}
\usepackage{graphics,color}
\usepackage{verbatim}
\usepackage{enumitem}

\usepackage{url}
\usepackage{dblfloatfix}
\usepackage{mathptmx}
\usepackage{pgf,tikz}
\usetikzlibrary{trees,automata,arrows,shapes.geometric}
\usepackage{xfrac}
\usepackage{amsfonts}
\usepackage{stmaryrd}
\usepackage{algorithm}
\usepackage[noend]{algpseudocode}
\usepackage{url}
\usepackage{colortbl}
\newtheorem{proposition}{Proposition}
\newtheorem{theorem}[proposition]{Theorem}
\newtheorem{corollary}[proposition]{Corollary}
\newtheorem{lemma}[proposition]{Lemma}

\newtheorem{example}{Example}

\newtheorem{remark}{Remark}
\newcommand*{\QEDB}{\hfill\ensuremath{\square}}

\newcommand{\Mod}[1]{\ (\textup{mod}\ #1)}
\newcommand{\etal}{\emph{et al.~}}
\newcommand{\eg}{\emph{e.g.}}
\newcommand{\ie}{\emph{i.e.}}

\def\wt{{\rm{wt}}}
\def\F{{\mathbb{F}}}

\DeclareMathOperator{\lcm}{lcm}

\algdef{SE}[DOWHILE]{Do}{doWhile}{\algorithmicdo}[1]{\algorithmicwhile\ #1}

\begin{document}
	\title{New Successor Rules to Efficiently Produce Exponentially Many Binary de Bruijn Sequences}
	
	\author{Zuling Chang, Martianus Frederic Ezerman, Pinhui Ke, and Qiang Wang
		\thanks{Z.~Chang is with the School of Mathematics and Statistics, Zhengzhou University, 450001 Zhengzhou, China. He is also with State Key Laboratory of Information Security, Institute of Information Engineering, Chinese Academy of Sciences, 100093 Beijing, China, email: zuling\_chang@zzu.edu.cn.}
		\thanks{M.~F.~Ezerman is with the School of Physical and Mathematical Sciences, Nanyang Technological University, 21 Nanyang Link, Singapore 637371, email: fredezerman@ntu.edu.sg.}
		\thanks{P.~Ke is with the School of Mathematics and Statistics, Fujian Normal University, 350117 Fuzhou, China, email: keph@fjnu.edu.cn.}
		\thanks{Q.~Wang is with the School of Mathematics and Statistics, Carleton University, 1125 Colonel By Drive, Ottawa, ON K1S 5B6, Canada, email: wang@math.carleton.ca}
		\thanks{The work of Z.~Chang is supported by the National Natural Science Foundation of China under Grant 61772476. Nanyang Technological University Grant 04INS000047C230GRT01 supports the research carried out by M.~F.~Ezerman. National Natural Science Foundation of China under Grant 61772292 helps P.~Ke with his research. Q.~Wang is supported by an individual grant from the Natural Sciences and Engineering Research Council of Canada.}
		\thanks{This work has been submitted to the IEEE for possible publication. Copyright may be transferred without notice, after which this version may no longer be accessible.}
	}
	\maketitle
	
	\begin{abstract}
		We put forward new general criteria to design successor rules that generate binary de Bruijn sequences. Prior fast algorithms based on successor rules in the literature are then shown to be special instances. We implemented the criteria to join the cycles generated by a number of simple feedback shift registers (FSRs) of order $n$. These include the pure cycling register (PCR) and the pure summing register (PSR). For the PCR,
		we define a transitive relation on its cycles, based on their weights. We also extend the choices of conjugate states by using shift operations. For the PSR, we define three distinct transitive relations on its cycles, namely a run order, a necklace order, and a mixed order. Using the new orders, we propose numerous classes of successor rules. Each class efficiently generates a number, exponential in $n$, of binary de Bruijn sequences. Producing the next bit in each such sequence takes $O(n)$ memory and $O(n)$ time. We implemented computational routines to confirm the claims.
	\end{abstract}
	
	
	\section{Introduction}\label{sec:intro}
	
	A $2^n$-periodic binary sequence is a {\it binary de Bruijn sequence} of order $n$ if every binary $n$-tuple occurs exactly once within each period. There are $2^{2^{n-1}-n}$ such sequences~\cite{Bruijn46}. They appear in many guises, drawing the attention of researchers from varied backgrounds and interests. Attractive qualities that include being balanced and having maximum period~\cite{Chan82,Golomb} make these sequences applicable in coding and communication systems. A subclass with properly calibrated nonlinearity property, while satisfying other measures of complexity, can also be useful in cryptography.
	
	Experts have been using tools from diverse branches of mathematics to study their generations and properties, see, \eg, the surveys in~\cite{Ral82} and~\cite{Fred82} for further details. Of enduring special interest are of course methods that excel in three measures: fast, with low memory requirement, and capable of generating a large number of sequences. Known constructions come with some trade-offs with respect to these measures. Notable examples include Lempel's $D$-Morphism~\cite{Lempel70}, an approach via preference functions described in~\cite{Alh12} and in~\cite{Golomb}, greedy algorithms with specific preferences, \eg, in~\cite{Martin34} and, more recently, in \cite{Chang20}, as well as various fast generation proposals, \eg, those in~\cite{Fred72} and in \cite{Dragon18}.
	
	The most popular construction approach is the {\it cycle joining method} (CJM)~\cite{Golomb}. It serves as the foundation of many techniques. A main drawback of the CJM, in its most general form,  is the amount of computation to be done prior to actually generating the sequences. Given a feedback shift register, one must first determine its cycle structure before finding the conjugate pairs to build the so-called adjacency graph. Enumerating the spanning trees comes next. Once these general and involved steps have been properly done, then generating a sequence, either randomly or based on a predetermined rule, is very efficient in both time and memory. The main advantage, if carried out in full, is the large number of output sequences, as illustrated in~\cite[Table 3]{Chang2019}.
	
	There are fast algorithms that can be seen as applications of the CJM on specially chosen conjugate pairs and designated initial states. They often produce a very limited number of de Bruijn sequences. One can generate a de Bruijn sequence, named the {\tt granddady} in~\cite{Fred72}, in $O(n)$ time and $O(n)$ space per bit. A related de Bruijn sequence, named the {\tt grandmama}, was built in~\cite{Dragon18}. Huang gave another early construction that joins the cycles of the {\it complementing circulating register} (CCR) in~\cite{Huang90}. Etzion and Lempel proposed some algorithms to generate de Bruijn sequences based on the {\it pure cycling register} (PCR) and the {\it pure summing register} (PSR) in~\cite{Etzion84}. Their algorithms generate a number, exponential in $n$, of sequences at the expense of higher memory requirement. 
	
	Jansen, Franx, and Boekee established a requirement to determine some conjugate pairs in \cite{Jansen91}, leading to another fast algorithm. In \cite{Sawada16}, Sawada, Williams, and Wong proposed a simple de Bruijn sequence construction, which is in fact a special case of the method in~\cite{Jansen91}. Gabric \etal generalized the last two works to form simple successor rule frameworks that yield more de Bruijn sequences in~\cite{Gabric18}. Further generalization to the constructions of $k$-ary de Bruijn sequences in~\cite{Sawada17} and~\cite{Gabric19} followed. Zhu \etal very recently in \cite{Zhu2021} proposed two efficient generic successor rules based on the properties of the feedback function $f(x_0, x_1, \ldots, x_{n-1}) = x_0 + x_1 + x_{n-1}$ for $n \geq 3$. Each rule produces at least $2^{n-3}$ binary de Bruijn sequences. They built upon the framework proposed in \cite{Gabric18}.
	
	\smallskip
	
	\noindent
	{\bf Our Contributions}
	\begin{enumerate}
		\item Paying close attention to the approach in~\cite{Jansen91} and the series of works that lead to the recently presented framework in~\cite{Gabric19}, we propose to generate de Bruijn sequences by using novel relations and orders on the cycles in combination with suitable successor rules,
		
		\item We define new classes of successor rules and, then, prove that they generate, respectively, a number, exponential in $n$, of de Bruijn sequences. In particular, the number of generated sequences based on the PCR of order $n$ is
		\[
		2(n-1)(n-2) \ldots 1 = 2 \cdot (n-1)!
		\]
		The cost to output the next bit is $O(n)$ time and $O(n)$ space. Nearly all known successor rules in the literature generate only a handful of de Bruijn sequences each. The few previously available approaches that can generate an exponential number of de Bruijn sequences require more space than the ones we are proposing.
		
		\item We implemented the criteria on some simple FSRs, especially on the PCR and the PSR of order $n$. Based on the properties of their respective cycles, we define several relations. For the PCR we order the cycles by their weights and the states in the respective cycles by their positions relative to the necklaces. On the cycles produced by the PSR we define a run order, a necklace order, and a mixed order that combines the weight order and the necklace order. 
		
		Using the new relations, we design numerous successor rules to efficiently generate de Bruijn sequences. The exact number of output sequences can be determined for many classes of the rules.
		Given a current state, in most occasions, the next bit takes only $O(n)$ space and $O(n)$ time to generate. In a few other instances, the process can be made even faster. We also demonstrate the explicit derivation of the feedback functions of some of the resulting sequences.
		
		\item Our results extend beyond providing a general formulation for already known fast algorithms that generate de Bruijn sequences by way of successor rules. The approach applies to \emph{any} FSR. To remain efficient, one should focus on classes of FSRs whose cycles have periods which are linear in $n$. There are plenty of such FSRs around for further explorations.
	\end{enumerate}
	
	A high level explanation of our approach is as follows. We begin with the set of cycles produced by any nonsingular feedback shift register. To join all of these cycles into a single cycle, \ie, to obtain a binary de Bruijn sequence, one needs to come up with a valid successor rule that assigns a uniquely identified state in one cycle to a uniquely identified state in another cycle and ensure that all of the cycles are joined in the end. If the cycles are represented by the vertices of an adjacency graph, then producing a de Bruijn sequence in the CJM corresponds to finding a spanning tree in the graph. The directed edges induced by a successor rule guide the actual process of generating the sequence. To certify that a successor rule can indeed yield a de Bruijn sequence we propose several new relations and orders on both the cycles and on the states in each cycle. These ensure the existence of spanning trees in the corresponding adjacency graphs. The relations and orders on the states are carefully chosen to guarantee that the next bit can be produced efficiently.
	
	We collect preliminary notions and several useful known results in Section~\ref{sec:prelim}. We present a new general criteria in Section~\ref{sec:general}. Section~\ref{sec:PCR} shows how to apply the criteria on the cycles of the PCR, leading to scores of new successor rules to generate de Bruijn sequences. Section~\ref{sec:PSR} gives a similar treatment on the PSR. The last section concludes this work by summarizing the contributions and listing some future directions.
	
	\section{Preliminaries}\label{sec:prelim}
	
	\subsection{Basic Definitions}
	
	An {\it $n$-stage shift register} is a circuit of $n$ consecutive storage units, each containing a bit. The circuit is clock-regulated, shifting the bit in each unit to the next stage as the clock pulses. A shift register generates a binary code if one adds a feedback loop that outputs a new bit $s_n$ based on the $n$ bits $\mathbf{s}_0= s_0,\ldots,s_{n-1}$, called an {\it initial state} of the register. The corresponding Boolean {\it feedback function} $f(x_0,\ldots,x_{n-1})$ outputs $s_n$ on input $\mathbf{s}_0$. A {\it feedback shift register} (FSR) outputs a binary sequence $\mathbf{s}=\{s_i\}=s_0,s_1,\ldots,s_n,$ $\ldots$ that satisfies the recursive relation
	\[
	s_{n+\ell} = f(s_{\ell},s_{\ell+1},\ldots,s_{\ell+n-1}) \text{ for } \ell = 0,1,2,\ldots.
	\]
	For $N \in \mathbb{N}$, if $s_{i+N}=s_i$ for all $i \geq 0$, then $\mathbf{s}$ is {\it $N$-periodic}
	or {\it with period $N$} and one writes $\mathbf{s}= (s_0,s_1,s_2,\ldots,s_{N-1})$. The least among all periods of $\mathbf{s}$ is called the {\it least period} of $\mathbf{s}$.
	
	We say that $\mathbf{s}_i= s_i,s_{i+1},\ldots,s_{i+n-1}$ is {\it the $i^{\rm th}$ state} of $\mathbf{s}$. Its {\it predecessor} is $\mathbf{s}_{i-1}$ while its {\it successor} is $\mathbf{s}_{i+1}$. For $s \in \F_2$, let $\bar{s} :=s+1 \in \F_2$. Extending the definition to any binary vector or sequence $\mathbf{s}= s_0,s_1,\ldots,s_{n-1}, \ldots$, let
	$\overline{\mathbf{s}} :=
	\overline{s_0},\overline{s_1},\ldots,\overline{s_{n-1}}, \ldots$. An arbitrary state $\mathbf{v}=v_0,v_1,\ldots,v_{n-1}$ of $\mathbf{s}$ has
	\[
	\widehat{\mathbf{v}}:=\overline{v_0},v_1,\ldots,v_{n-1} \mbox{ and }
	\widetilde{\mathbf{v}}:=v_0,\ldots,v_{n-2},\overline{v_{n-1}}
	\]
	as its {\it conjugate} state and {\it companion} state, respectively. Hence, $(\mathbf{v}, \widehat{\mathbf{v}})$ is a {\it conjugate pair} and $(\mathbf{v}, \widetilde{\mathbf{v}})$ is a {\it companion pair}.
	
	For any FSR, distinct initial states generate distinct sequences. There are $2^n$ distinct sequences generated from an FSR with feedback function $f(x_0,x_1,\ldots,x_{n-1})$. All these sequences are periodic if and only if $f$ is {\it nonsingular}, \ie, $f$ can be written as
	\[
	f(x_0,x_1,\ldots,x_{n-1})=x_0+h(x_1,\ldots,x_{n-1}),
	\]
	for some Boolean function $h(x_1,\ldots,x_{n-1})$ whose domain is $\F_2^{n-1}$~\cite[p.~116]{Golomb}. All feedback functions in this paper are nonsingular. An FSR is {\it linear} or an LFSR if its feedback function has the form
	\[
	f(x_0,x_1,\ldots,x_{n-1}) = x_0 + c_1 x_1 + \ldots + c_{n-1} x_{n-1}\mbox{, with }\ c_i \in \F_2,
	\]
	and its {\it characteristic polynomial} is 
	\[
	f(x)=x^n+c_{n-1}x^{n-1}+\cdots+c_1x+1\in \mathbb{F}_2[x].
	\]
	Otherwise, it is {\it nonlinear} or an NLFSR. Further properties of LFSRs are treated in, \eg,~\cite{GG05} and~\cite{LN97}.
	
	For an $N$-periodic sequence $\mathbf{s}$, the {\it left shift operator} $L$ maps $(s_0,s_1,\ldots,s_{N-1}) \mapsto (s_1,s_2,\ldots,s_{N-1},s_0)$, with the convention that $L^0$ fixes $\mathbf{s}$. The {\it right shift operator} $R$ is defined analogously. The set
	\begin{equation}\label{eq:cycle}
		[\mathbf{s}]:=\left\{\mathbf{s},L\mathbf{s},\ldots,L^{N-1}\mathbf{s} \right\} =
		\left\{\mathbf{s},R\mathbf{s},\ldots,R^{N-1}\mathbf{s} \right\}
	\end{equation}
	is a {\it shift equivalent class}. Sequences in the same shift equivalent class correspond to the same cycle in the state diagram of FSR \cite{GG05}.
	We call a periodic sequence in a shift equivalent class a {\it cycle}. If an FSR with feedback function $f$ generates $r$ disjoint cycles
	$C_1, C_2, \ldots, C_r$, then its {\it cycle structure} is
	\[
	\Omega(f)=\{C_1, C_2, \ldots, C_r\}.
	\]
	A cycle can also be viewed as a set of consecutive $n$-stage states in the corresponding periodic sequence. Since the cycles are disjoint, we can write
	\[
	\mathbb{F}_2^n=C_1 \cup C_2 \cup \ldots \cup C_r.
	\]
	When $r=1$, the corresponding FSR is of {\it maximal length} and its output is a de Bruijn sequence of order $n$.
	
	The {\it weight} of an $N$-periodic cycle $C$, denoted by $\wt(C)$, is
	\[
	|\{ 0 \leq j \leq N-1 : c_j = 1  \}|.
	\]
	Similarly, the weight of a state is the number of $1$s in the state. The lexicographically least $N$-stage state in any $N$-periodic cycle is called its {\it necklace}. As discussed in, \eg, \cite{Booth80} and~\cite{Gabric18}, there is a fast algorithm that determines whether or not a state is a necklace in $O(N)$ time. In fact, one can efficiently sort all distinct states in $C$. The standard {\tt python} implementation is {\tt timsort}~\cite{timsort}. It was developed by Tim Peters based on McIlroy's techniques in~\cite{McIlroy93}. In the worst case, its space and time complexities are $O(N)$ and $O(N \, \log N)$ respectively. A closely related proposal, by Buss and Knop, is in~\cite{Buss2019}.
	
	Given disjoint cycles $C$ and $C'$ in $\Omega(f)$ with the property that
	some state $\mathbf{v}$ in $C$ has its conjugate state $\widehat{\mathbf{v}}$ in $C'$, interchanging the successors of $\mathbf{v}$ and $\widehat{\mathbf{v}}$ joins $C$ and $C'$ into a cycle whose feedback function is
	\begin{equation}\label{eq:newfeedback}
		\widehat{f}:=f(x_0,x_1,\ldots,x_{n-1})+\prod_{i=1}^{n-1}(x_i+\overline{v_{i}}).
	\end{equation}
	Similarly, if the companion states $\mathbf{v}$ and $\widetilde{\mathbf{v}}$ are in two distinct cycles, then interchanging their predecessors joins the two cycles. If this process can be continued until all cycles that form $\Omega(f)$ merge into a single cycle, then we obtain a de Bruijn sequence. The CJM is, therefore, predicated upon knowing the cycle structure of $\Omega(f)$ and is closely related to a graph associated to the FSR.
	
	Given an FSR with feedback function $f$, its {\it adjacency graph} $G_f$, or simply $G$ if $f$ is clear, is an undirected multigraph whose vertices correspond to the cycles of $\Omega(f)$. The number of edges between two vertices is the number of shared conjugate (or companion) pairs, with each edge labelled by a specific pair. It is well-known that there is a bijection between the set of spanning trees of $G$ and the set of all inequivalent de Bruijn sequences constructible by the CJM on input $f$.
	
	We state the Generalized Chinese Remainder Theorem,  which will be used as an enumeration tool in Section~\ref{sec:PSR}.
	
	\begin{theorem}\cite[Section 2.4]{Ding96}\label{thm:gcrt}
		Let $m_1,\ldots,m_k$ be positive integers. For a set of integers $a_1,\ldots,a_k$, the system of congruences
		\[
		\{ x \equiv a_i \Mod{m_i} \mbox{ for all } i \in \{1,\ldots,k \} \}
		\]
		is solvable if and only if
		\begin{equation}\label{equ:gcrt}
			a_i \equiv a_j \Mod{\gcd(m_i,m_j)} \mbox{ for all } 
			1 \leq i \neq j \leq k.
		\end{equation}
		If the equivalence in (\ref{equ:gcrt}) holds, then the solution is unique modulo $\lcm(m_1,\ldots,m_k)$.
	\end{theorem}
	
	\subsection{Properties of Some Feedback Shift Registers}
	
	We now introduce some simple FSRs to be used later.
	
	A {\it pure cycling register} (PCR) {\it of order} $n$ is an LFSR with feedback function and characteristic polynomial
	\begin{equation}\label{eq:PCR}
		f_{\rm PCR}(x_0,x_1,\ldots,x_{n-1})= x_0 \mbox{ and } f_{\rm PCR}(x)=x^n+1.
	\end{equation}
	Let $\phi(\cdot)$ be the Euler totient function. The number of distinct cycles in $\Omega(f_{\rm PCR})$ is known, \eg, from~\cite{Golomb}, to be
	\begin{equation}\label{eq:Zn}
		Z_n:=\frac{1}{n}\sum_{d|n} \phi(d) 2^{\frac{n}{d}}.
	\end{equation}
	By definition, all states in any given $n$-periodic cycle $C_{\rm PCR}:=(c_0,c_1,\ldots,c_{n-1}) \in \Omega(f_{\rm PCR})$ have the same number of ones.
	
	A {\it pure summing register} (PSR) {\it of order} $n$ is an LFSR with feedback function and characteristic polynomial
	\begin{equation}\label{eq:PSR}
		f_{\rm PSR}(x_0,x_1,\ldots,x_{n-1})= \sum_{j=0}^{n-1} x_j \mbox{ and }
		f_{\rm PSR} (x) = \sum_{j=0}^n x^j.
	\end{equation}
	The cycles of the PSR share some interesting properties. If $C_{\rm PSR}$ is any cycle generated by the PSR of order $n$, then its least period divides $n+1$. Hence, we can write it as an $(n+1)$-periodic cycle, \ie,
	$C_{\rm PSR}:=(c_0,c_1,\ldots,c_{n})$. Notice that $\wt\left(C_{\rm PSR}\right)$ must be even. Let $n\geq2$. If $n$ is odd, we can write $n := 2^tn'-1$. The number of distinct cycles in $\Omega(f_{\rm PSR})$ is 
	\begin{equation}\label{eq:psrnum}
		Z_{n+1} - \frac{1}{2(n+1)}
		\left( \sum\limits_{d \mid n'}\phi \left(\frac{n'}{d}\right) 2^{d2^t}\right),
	\end{equation}
	where $Z_{n+1}$ is computed based on (\ref{eq:Zn}). The number in (\ref{eq:psrnum}) simplifies to $\frac{1}{2}\, Z_{n+1}$ if $n$ is even.
	
	The {\it complemented PSR}, also known as the CSR, of order $n$ is an LFSR with feedback function
	\begin{equation}\label{eq:CSR}
		f_{\rm CSR}(x_0,x_1,\ldots,x_{n-1})= 1 + \sum_{j=0}^{n-1} x_j.
	\end{equation}
	It assigns the next bit to be the complement of the bit produced by the feedback function $f_{\rm PSR}$ in (\ref{eq:PSR}), when given the same input. Hence, the least period of any cycle $C_{\rm CSR}$ divides $n+1$ and the weight of any $C_{\rm CSR}:=(e_0,e_1,\ldots,e_n)$ is odd.
	
	For a fixed $n$, the PSR and the CSR have analogous properties that can be easily inferred from each others. In what follows, our focus is on the PSR since the corresponding results on the CSR become immediately apparent with the proper adjustment.
	
	\subsection{Jansen-Franx-Boekee (JFB) Algorithm}\label{JFB}
	
	In \cite{Jansen91}, Jansen \etal proposed an algorithm to generate de Bruijn sequences by the CJM. Suppose that the FSR with a feedback function $f(x_0,x_1,\ldots,x_{n-1})$ is given.
	They defined the {\it cycle representative} of any cycle of the FSR to be its lexicographically smallest $n$-stage state. If the FSR is the PCR of order $n$, then it is clear that the cycle representative is its necklace. Based on the cycle representative, we can impose an order on the cycles. For arbitrary cycles $C$ and $C'$ in $\Omega_f$, we say that $C { \prec}_{\rm lex} C'$ if and only if the cycle representative of $C$ is lexicographically less than that of $C'$. This {\it lexicographic order} defines a total order on the cycles of the said PCR.
	
	On current state ${\bf s}_i=s_i,s_{i+1},\ldots,s_{i+n-1}$, the next state ${\bf s}_{i+1}=s_{i+1},s_{i+2},\ldots,s_{i+n}$ is produced based on the assignment rule in Algorithm~\ref{alg:Jansen}. The correctness of the JFB Algorithm rests on the fact that the cycle representative in any cycle $C_1$ which does not contain the all zero state $0,\ldots,0$ is unique. Its companion state is guaranteed to be in another cycle $C_2$ with $C_2 {\prec}_{\rm lex} C_1$. This ensures that we have a spanning tree and, hence, the resulting sequence must be de Bruijn. 
	
	\begin{algorithm}[h!]
		\caption{{Jansen-Franx-Boekee (JFB) Algorithm}}
		\label{alg:Jansen}
		\begin{algorithmic}[1]
			\If{${\bf s}_i=s_i,0,\ldots,0$}
			\State{${\bf s}_{i+1} \gets 0,\ldots,0,s_i+1$}
			\Else
			\If{$s_{i+1},\ldots,s_{i+n-1},0$ or $s_{i+1},\ldots,s_{i+n-1},1$ is a cycle representative}
			\State{${\bf s}_{i+1} \gets s_{i+1},\ldots,s_{i+n-1},f(s_i,\ldots,s_{i+n-1})+1$}
			\Else
			\State{${\bf s}_{i+1} \gets s_{i+1},\ldots,s_{i+n-1},f(s_i,\ldots,s_{i+n-1})$}
			\EndIf
			\EndIf
		\end{algorithmic}
	\end{algorithm}
	
	The main task of keeping track of the cycle representatives in Algorithm~\ref{alg:Jansen} may require a lot of time if the least periods of the cycles are large. For cases where all cycles produced by a given FSR have small least periods, \eg, in the case of the PCR or the PSR of order $n$, the algorithm generates de Bruijn sequences very efficiently. The space complexity is $O(n)$ and the time complexity lies between $O(n)$ and $O(n\log n)$ to output the next bit.
	
	Sawada \etal proposed a simple fast algorithm on the PCR to generate a de Buijn sequence \cite{Sawada16}. Their algorithm is a special case of the JFB Algorithm. Later, in \cite{Gabric18}, Gabric and the authors of~\cite{Sawada16} considered the PCR and the complemented PCR, also known as the CCR, and proposed several fast algorithms to generate de Bruijn sequences by ordering the cycles lexicographically according to their respective necklace and co-necklace. They replace the generating algorithm by some {\it successor rule}. 
	
	The general thinking behind the approach is as follows. Given an FSR with a feedback function $f(x_0,x_1,\ldots,x_{n-1})$, let $A$ label some condition which guarantees that the resulting sequence is de Bruijn. For any state $\mathbf{c}:=c_0,c_1,\ldots,c_{n-1}$, the successor rule assigns
	\begin{equation}\label{equ:1}
		\rho_{A}(\mathbf{c})=
		\begin{cases}
			f(\mathbf{c})+1 \mbox{, if }
			\mathbf{c} \mbox{ satisfies } A,\\
			f(\mathbf{c}) \mbox{, otherwise.}
		\end{cases}
	\end{equation}
	
	The usual successor of $\mathbf{c}$ is $c_1,\ldots,c_{n-1}, f(c_0,\ldots,c_{n-1})$. Every time $\mathbf{c}$ satisfies Condition $A$, however, its successor is \emph{redefined} to be $c_1,\ldots,c_{n-1}, f(c_0,\ldots,c_{n-1})+1$. The last bit of the successor is \emph{the complement} of the last bit of the usual successor under the feedback function $f$. The basic idea of a successor rule is to determine spanning trees in $G_f$ by identifying a suitable Condition $A$. Seen in this light, the rule implements the CJM by assigning successors to carefully selected states.
	
	We will devise numerous new successor rules to join the cycles produced by the PCR and the PSR of any order $n$. Known successor rules in the literature will subsequently be shown to be special instances of our more general results.
	
	\section{New General Criteria for Successor Rules}\label{sec:general}
	
	New successor rules for de Bruijn sequences can be established by defining some relations or orders on the cycles of FSRs with special properties to construct spanning trees in $G_f$. This section proves a general criteria that such rules must meet. The criteria will be applied successfully, in latter sections, to the PCR and the PSR of any order $n$. The generality of the criteria allows for further studies to be conducted on the feasibility of using broader families of FSRs for fast generation of de Bruijn sequences.
	
	We adopt set theoretic definitions and facts from \cite{Halmos}. Given $\Omega_f$, we define a binary {\it relation} $\prec$ on $\Omega_f:=\{C_1,C_2,\ldots,C_r\}$ as a set of ordered pairs in $\Omega_f$. If $C \prec C$ for every $C \in \Omega_f$, then $\prec$ is said to be {\it reflexive}. Let $1 \leq i,j,k \leq r$. We say that $\prec$ is {\it transitive } if $C_i \prec C_j$ and $C_j \prec C_k$, together, imply $C_i \prec C_k$. It is {\it symmetric} if $C_i \prec C_j$ implies $C_j \prec C_i$ and {\it antisymmetric} if the validity of both $C_i \prec C_j$ and $C_j \prec C_i$ implies $C_i = C_j$.
	
	The relation $\prec$ is called a {\it preorder} on $\Omega_f$ if it is reflexive and transitive. It becomes a {\it partial order} if it is an antisymmetric preorder. If $\prec$ is a partial order with either $C_i \prec C_j$ or $C_j \prec C_i$, for any $C_i$ and $C_j$, then it is a {\it total order}. A totally ordered set $\Omega_f$ is called a {\it chain}. Hence, we can now say that $\prec_{\rm lex}$ defined in Subsection~\ref{JFB} is a total order on the corresponding chain $\Omega_f$.
	
	\begin{theorem}\label{thm:order}
		Given an FSR with feedback function $f$, let $\prec$ be a transitive relation on $\Omega(f):=\{C_1,C_2,\ldots,C_r\}$ and let $1 \leq i, j \leq r$.
		\begin{enumerate}
			\item Let there be a unique cycle $C$ with the property that $C \prec C'$ for any cycle $C' \neq C$, \ie, $C$ is the {\it unique smallest cycle} in $\Omega(f)$. Let $\rho$ be a successor rule that can be well-defined as follows. If any cycle $C_i \neq C$ contains a {\it uniquely defined state} whose successor can be assigned by $\rho$ to be a state in a cycle $C_j \neq C_i$ with $C_j \prec C_i$, then $\rho$ generates a de Bruijn sequence.
			\item Let there be a unique cycle $C$ with the property that $C' \prec C$ for any cycle $C' \neq C$, \ie, $C$ is the {\it unique largest cycle} in $\Omega(f)$. Let $\rho$ be a successor rule that can be well-defined as follows. If any cycle $C_i \neq C$ contains a {\it uniquely defined state} whose successor can be assigned by $\rho$ to be a state in a cycle $C_j \neq C_i$ with $C_i \prec C_j$, then $\rho$ generates a de Bruijn sequence.
		\end{enumerate}
	\end{theorem}
	
	\begin{IEEEproof}
		We prove the first case by constructing a rooted tree whose vertices are all of the cycles in $\Omega(f)$. This exhibits a spanning tree in the adjacency graph of the FSR according to the specified successor rule. The second case can be similarly argued.
		
		Based on the condition set out in the first case, each $C_i \neq C$ contains a unique state whose assigned successor under $\rho$ is in $C_j \neq C_i$, revealing that $C_i$ and $C_j$ are adjacent. Since $C_j \prec C_i$, we direct the edge {\bf from} $C_i$ {\bf to} $C_j$. It is easy to check that, except for $C$ whose outdegree is $0$, each vertex has outdegree $1$. Since $\prec$ is transitive, there is a unique path from the vertex to $C$. We have thus built a spanning tree rooted at $C$.
	\end{IEEEproof}
	
	\begin{remark}\label{rem:moregeneral} 
		Armed with Theorem~\ref{thm:order}, one easily verifies that the JFB Algorithm and the successor rules proposed in~\cite{Gabric18} are valid. In both references, the relation is a total order. In our present notation, \cite[Theorem 3.5]{Gabric18} says that a successor rule generates a de Bruijn sequence if $\Omega_f$ is a chain. The cycles, possibly with a relabelling of the indices, can be presented as $C_1 \prec C_2 \prec \ldots \prec C_r$. In each $C_i$, with $1 <i \leq r$, there exists a unique state whose successor can be defined to be a state in $C_{j}$ with $j < i$. 
		
		Known successor rules in the literature have so far been mostly based on the lexicographic order in $\Omega_f$ for a chosen $f$. A notable exception is the class of successor rules in \cite[Section 4]{Zhu2021}. As we will soon see, many relations in the set of cycles that we are defining later do not constitute total orders, so \cite[Theorem 3.5]{Gabric18} cannot be used to prove the correctness of the resulting successor rules directly. 
		Theorem~\ref{thm:order} relaxes the requirement and works as long as the relation is transitive. In the sequel we show that many alternatives to $\prec_{\rm lex}$ can be devised to efficiently generate de Bruijn sequences. The corresponding successor rules are simple to state and straightforward to validate. Thus, Theorem~\ref{thm:order} can be viewed as the generalization of \cite[Theorem 3.5]{Gabric18}. \QEDB
	\end{remark}	
	
	There are two tasks to carry out in using Theorem~\ref{thm:order}. First, one must define a suitable transitive relation among the cycles to obtain the unique smallest or largest cycle $C$. The second task is to determine the unique state in each cycle. A sensible approach is to designate a state $\mathbf{v}$ as the {\it benchmark state} in each cycle $C$. We then uniquely define a state $\mathbf{w}$ in $C$ with respect to the benchmark state. The cycle representative, \ie, the necklace in the PCR, is the most popular choice for $\mathbf{v}$. In this paper we mainly use the necklace as the benchmark state in each cycle.
	
	The next two sections examine some FSRs whose cycles have small respective least periods. Based on the properties of their respective cycle structures, we define several relations or orders to come up with new successor rules that meet the criteria in Theorem~\ref{thm:order}.
	
	\section{Successor Rules from Pure Cycling Registers}\label{sec:PCR}
	
	This section applies the criteria in Theorem \ref{thm:order} to the PCR of any order $n$. A good strategy is to consider the positions of the states in each cycle {\it relative to its necklace} by ordering the states in several distinct manners. This general route is chosen since
	we can check whether or not a state is a necklace in $O(n)$ time and $O(n)$ space. If the relative position of a state to the necklace is efficient to pinpoint, then the derived successor rule also runs efficiently. 
	
	\subsection{The Weight Relation on the Pure Cycling Register}\label{subsec:weightPCR}
	
	The cycles of the PCR share a nice property. All of the states in any cycle $C$ are shift-equivalent and share the same weight $\wt(C)$. Hence, we can define a {\it weight relation} on the cycles based simply on their respective weights. For cycles $C_i \neq C_j$, we say that $C_i \prec_{\wt} C_j$ if and only if $\wt(C_i) < \wt(C_j)$.
	
	The relation $\prec_{\wt}$ is not even a preorder, making it differs qualitatively from the lexicographic order.
	
	\begin{example}
		The PCR of order $6$ generates $C_1:=(001001)$ and $C_2:=(000111)$. Lexicographically $C_1 \succ_{\rm lex} C_2$ because the necklace $001001$ in $C_1$ is lexicographically larger than the necklace $000111$ in $C_2$. In the weight relation, however, $C_1 \prec_{\wt}  C_2$ since $\wt(C_1 ) = 2 < 3 = \wt(C_2)$. \QEDB
	\end{example}
	
	The following successor rules rely on the weight relation.
	
	\begin{theorem}\label{thm1}
		For the PCR of order $n$, if a successor rule $\rho(x_0,x_1,\ldots,x_{n-1})$ satisfies one of the following conditions, then it generates a de Bruijn sequence.
		\begin{enumerate}[nolistsep]
			\item For any $C_i \neq (0)$, the rule $\rho$ exchanges the successor of a {\it uniquely determined state} $\mathbf{v}_i \in C_i$ with a state $\mathbf{w}_j$ in $C_j$, where $C_j \prec_{\wt}  C_i$.
			\item For any $C_i \neq (1)$, the rule $\rho$ exchanges the successor of a {\it uniquely determined state} $\mathbf{v}_i \in C_i$ with a state $\mathbf{w}_j$ in $C_j$, where $C_i \prec_{\wt}  C_j$.
		\end{enumerate}
	\end{theorem}
	
	\begin{IEEEproof}
		To prove the first case, note that $(0) \prec_{\wt} C_i$ for any $C_i \neq (0)$ in
		$\Omega(f_{\rm PCR})$. By the stated condition, $C_i$ contains a unique state $\mathbf{v}_i$ such that its conjugate $\mathbf{w}_j := \widehat{\mathbf{v}_i}$ is in $C_j$ and $\wt(C_j) < \wt(C_i) $. The successor rule $\rho$ satisfies the criteria in Theorem~\ref{thm:order}. The proof for the second case is similar.
	\end{IEEEproof}
	
	Theorem~\ref{thm1} reduces the task to generate de Bruijn sequences by using $\rho$ to performing one of two procedures. The first option is to find the {\it uniquely determined state} $\mathbf{v}_i \in C_i \neq (0)$ whose conjugate state $\widehat{\mathbf{v}_i}$ is guaranteed to be in $C_j$ with $\wt(C_j) < \wt(C_i)$. The second option is to find the {\it uniquely determined state} $\mathbf{v}_i$ in each $C_j \neq (1)$ whose conjugate state $\widehat{\mathbf{v}_i}$ is guaranteed to be in $C_j$ with $\wt(C_j) > \wt(C_i)$. If, for every $C_i$, its $\mathbf{v}_i$ can be determined quickly, then generating the de Bruijn sequence is efficient. Following the two cases in Theorem \ref{thm1}, the rule $\rho$ comes in two forms. Let $\mathbf{c}:=c_0,c_1,\ldots,c_{n-1}$.
	
	First, let $\mathcal{A}$ be
	\begin{quote}
		In $C:=(0,c_1,\ldots,c_{n-1})$, the uniquely determined state $\mathbf{v}$ is $0,c_1,\ldots,c_{n-1}$.  Its conjugate $\widehat{\mathbf{v}}$ has $\wt(\widehat{\mathbf{v}}) > \wt(\mathbf{v})$, which implies $\widehat{\mathbf{v}}$ is in $C'$ with $C \prec_{\wt} C'$.
	\end{quote}
	It is then straightforward to confirm that the relevant requirement in Theorem \ref{thm1} is met by
	\begin{equation}\label{equ:2}
		\rho_{\mathcal{A}}(\mathbf{c}) =
		\begin{cases}
			\overline{c_0} \mbox{, if } 0,c_1,\ldots,c_{n-1} \mbox{ satisfies } \mathcal{A},\\
			c_0 \mbox{, otherwise}.
		\end{cases}
	\end{equation}
	
	Second, let $\mathcal{B}$ be
	\begin{quote}
		In $C:=(c_1,\ldots,c_{n-1},1)$, the uniquely determined state $\mathbf{v}$ is  $c_1,\ldots,c_{n-1},1$. Its companion $\widetilde{\mathbf{v}}$ has $\wt(\widetilde{\mathbf{v}}) < \wt(\mathbf{v})$, which means that $\widetilde{\mathbf{v}_i}$ is in $C'$ with $C' \prec_{\wt} C$.
	\end{quote}
	Hence, the successor rule
	\begin{equation}\label{equ:3}
		\rho_{\mathcal{B}}(\mathbf{c})=
		\begin{cases}
			\overline{c_0} \mbox{, if } c_1,\ldots,c_{n-1},1  \mbox{ satisfies } \mathcal{B},\\
			c_0 \mbox{, otherwise,}
		\end{cases}
	\end{equation}
	fulfills the requirement in Theorem \ref{thm1}.
	
	The next example provides respective explicit rules based on $\rho_{\mathcal{A}}$ in (\ref{equ:2}) and $\rho_{\mathcal{B}}$ in (\ref{equ:3}).
	
	\begin{example}\label{ex3}
		Let $\mathcal{A}$ in (\ref{equ:2}) be formulated as follows. We apply consecutive right shifts on $\mathbf{v} :=0,c_1,\ldots,c_{n-1}$ until we encounter, for the first time, a state whose first bit is $0$ and it is a necklace. It is $\mathbf{v}$ itself if the cycle contains only a single state whose first bit is $0$. Since the necklace is unique, the state $\mathbf{v}$ in the cycle $(\mathbf{v})\neq (1)$ is uniquely determined. By how the successor rule is designed, the successor of $\mathbf{v}$ is therefore $c_1,\ldots,c_{n-1},1$. Its weight is $\wt(\mathbf{v}) + 1$. The criteria of Theorem \ref{thm:order} is met. Hence, the generated sequence is de Bruijn.
		
		For $\mathbf{v}:=0,c_1,\ldots,c_{n-1}$, if there is an integer $1 \leq j < n$ such that $j$ is the largest index for which $c_j = 0$, then $\mathbf{u} := 0,c_{j+1},\ldots,c_{n-1},0,c_1,\ldots,c_{j-1}$. Otherwise, $\mathbf{u} :=\mathbf{v}$. The successor rule in the preceding paragraph simplifies to
		\begin{equation}\label{new1}
			\rho_{\mathcal{A}}(c_0,c_1,\ldots,c_{n-1}) =
			\begin{cases}
				\overline{c_0} \mbox{, if } \mathbf{u} \mbox{ is a necklace}, \\
				c_0 \mbox{, otherwise}.
			\end{cases}
		\end{equation}
		Using $\rho_{\mathcal{S}}$ in (\ref{new1}) and $n=6$, we can choose $c_0,c_1,\ldots,c_{n-1}$ as the uniquely determined state in each cycle $(c_0,c_1,\ldots,c_{n-1}) \neq (1)$. Table~\ref{table:ustate} lists the states to choose from according to the weight of their respective cycles. It is then easy to construct the spanning tree in Figure~\ref{fig:ex2}. The directed edge from $C_i$, for each $1 \leq i < 14$, is labelled by the pair $(\mathbf{v}_i,\widehat{\mathbf{v}_i})$. The resulting de Bruijn sequence is
		\begin{equation*}\label{dbs1}
			(00000010 \ 10111000 \ 11101100 \ 10010110 \ 11111100 \ 11110100 \ 11000011 \ 01010001).
		\end{equation*}
		This sequence is distinct from the output of any previously known successor rule in the literature.
		
		\begin{table}
			\caption{The uniquely determined states for $\rho$ in (\ref{new1}) when applied to the PCR of order $6$.}
			\label{table:ustate}
			\renewcommand{\arraystretch}{1.2}
			\centering
			\begin{tabular}{c c c | c c c}
				\hline
				$i$ & $\wt(C_i)$ & $\mathbf{v}_i \in C_i:=(\mathbf{v}_i)$ & $i$ & $\wt(C_i)$ & $\mathbf{v}_i \in C_i:=(\mathbf{v}_i)$\\
				\hline
				$1$ & $0$ & $000000$ & $8$ & $3$ & $010110$ \\
				$2$ & $1$ & $000010$ & $9$ & $3$ & $011010$ \\
				$3$ & $2$ & $000110$ & $10$ & $4$ & $011110$ \\
				$4$ & $2$ & $001010$ & $11$ & $4$ & $011101$ \\
				$5$ & $2$ & $010010$ & $12$ & $4$ & $011011$ \\
				$6$ & $3$ & $010101$ & $13$ & $5$ & $011111$ \\
				$7$ & $3$ & $001110$ &  & & \\
				\hline
			\end{tabular}
		\end{table}
		
		\begin{figure}
			\centering
			\begin{tikzpicture}
				[
				> = stealth,
				shorten > = 3pt,
				auto,
				node distance = 1.4cm,
				semithick
				]
				
				\tikzstyle{every state}=
				\node[rectangle,fill=white,draw,rounded corners,minimum size = 4mm]
				
				\node[state] (1) {$C_1$};
				\node[state] (2) [right of=1] {$C_2$};
				
				\node[state,fill=lightgray] (3) [right of=2] {$C_4$};
				\node[state,fill=lightgray] (4) [right of=3]{$C_3$};
				\node[state,fill=lightgray] (5) [right of=4] {$C_5$};
				
				\node[state,fill=blue!20] (6) [below of=3] {$C_6$};
				\node[state,fill=blue!20] (7) [right of=6] {$C_9$};
				\node[state,fill=blue!20] (8) [right of=7] {$C_7$};
				\node[state,fill=blue!20] (9) [right of=8] {$C_8$};
				
				\node[state,fill=red!40] (10) [below of=6] {$C_{10}$};
				\node[state,fill=red!40] (11) [right of=10] {$C_{11}$};
				\node[state,fill=red!40] (12) [right of=11] {$C_{12}$};
				
				\node[state] (13) [below of=11] {$C_{13}$};
				\node[state] (14) [right of=13] {$(1)$};
				
				\path[->] (1) edge (2);
				\path[->] (2) edge (3);
				\path[->] (3) edge (6);
				\path[->] (6) edge (11);
				\path[->] (11) edge (13);
				\path[->] (13) edge (14);
				
				\path[->] (4) edge (7);
				\path[->] (5) edge (9);
				\path[->] (7) edge(11);
				\path[->] (8) edge (11);
				\path[->] (9) edge (12);
				\path[->] (10) edge (13);
				\path[->] (12) edge (13);
				
			\end{tikzpicture}
			
			\caption{The spanning tree produced by the successor rule $\rho$ in (\ref{new1}) when applied to the PCR of order $6$. The cycles in gray, blue, and red are of the same weights $2$, $3$, and $4$, respectively.}
			\label{fig:ex2}
		\end{figure}
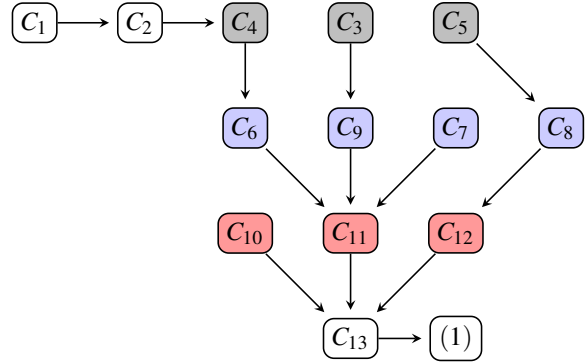
		
		Let us specify $\mathcal{A}$ in (\ref{equ:2}) to be
		\begin{quote}
			The state $\mathbf{v}$ is a necklace.
		\end{quote}
		We obtain the PCR$4$ de Bruijn sequence in \cite{Gabric18}. 
		
		Let $\mathcal{A}$ be 
		\begin{quote}
			$L^k \mathbf{v}$ is a necklace with $k$ being the smallest positive integer such that 
			$L^k \mathbf{v}$ has $0$ as its first bit.
		\end{quote}
		The resulting sequence is the {\tt granddaddy}.
		
		One can also formulate specific successor rules based on (\ref{equ:3}). Let $\mathbf{v}:=c_1,\ldots,c_{n-1},1$ be a state in 
		$C \neq (0)$. Let $\mathcal{B}$ be
		\begin{quote}
			$L^k \mathbf{v}$ is a necklace with $k$ being the smallest positive integer such that
			$L^k \, \mathbf{v}$ has $1$ as its last bit.
		\end{quote}
		If there is only one state whose last bit is $1$ in $C$, then $L^k \, \mathbf{v} = \mathbf{v}$. Since the necklace is unique, $\mathbf{v}$ in the cycle $(\mathbf{v})$ is uniquely determined. Let $\mathbf{w}$ be the predecessor of $\mathbf{v}$ under $f_{\rm PCR}$. The rule $\rho_{\mathcal{B}}$ assigns $c_1,\ldots,c_{n-1},0$, which has weight $\wt(\mathbf{v})-1$, as the successor of $\mathbf{w}$. Thus, by Theorem \ref{thm:order}, the generated sequence is de Bruijn.
		
		For $\mathbf{v}:=c_1,\ldots,c_{n-1},1$, if there is an integer $1 \leq j < n$ such that $j$ is the least index that satisfies $c_j=1$, then $\mathbf{u} :=c_{j+1},\ldots,c_{n-1},1,c_1,\ldots,c_{j-1},1$. Otherwise, $\mathbf{u} := \mathbf{v}$. The rule in the preceding paragraph becomes
		\begin{equation}\label{new2}
			\rho_{\mathcal{B}}(c_0,c_1,\ldots,c_{n-1}) =
			\begin{cases}
				\overline{c_0} \mbox{, if } \mathbf{u} \mbox{ is a necklace}, \\
				c_0 \mbox{, otherwise}.
			\end{cases}
		\end{equation}
		On the PCR of order $6$, the resulting sequence is
		\begin{equation*}\label{dbs2}
			(0000 0010 \ 0111 1001 \ 1010 0101 \ 1011 0010 \ 
			0011 1000 \ 1010 1111 \ 1101 1101 \ 0100 0011),
		\end{equation*}
		which is again distinct from any that can be produced based on previously known successor rule.
		
		Two more observations are worth mentioning. Let $\mathcal{B}$ be
		\begin{quote}
			The state $\mathbf{v}$ is a necklace.
		\end{quote}
		Then the output is the PCR$3$(J$1$) de Bruijn sequence in \cite{Gabric18}. We obtain the {\tt grandmama} when $\mathcal{B}$ is 
		\begin{quote}
			$R^k \mathbf{v}$ is a necklace with $k$ being the least positive integer such that $R^k \mathbf{v}$ has $1$ as its last bit.
		\end{quote}
		\QEDB
	\end{example}
	
	Based on $\mathcal{A}$ and $\mathcal{B}$, valid successor rules can be easily formulated once we manage to determine a unique state whose first bit is $0$, respectively, whose last bit is $1$, in each $C \neq (1)$, respectively, $C \neq (0)$. There are numerous ways to do so if one sets aside the issue of efficiency. Let us consider valid successor rules designed based only on (\ref{equ:2}) on the PCR. A direct inspection on the list of $C_i =(\mathbf{u}_i)$ in Table~\ref{table:ustate} confirms that for $n=6$ the number of resulting de Bruijn sequences is
	\[
	2^3 \cdot 3^3 \cdot 4^2 \cdot 5 = 17,280 \approx 2^{14}.
	\]
	When $n=7$, the number is
	\[
	2^3 \cdot 3^5 \cdot 4^5 \cdot 5^3 \cdot 6 = 1,492,992,000 \approx 2^{30.475}.
	\]

	\begin{table*}
		\caption{Examples of de Bruijn sequences from Proposition \ref{prop:sf0} with $n=6$.}
		\label{table:AtoE}
		\renewcommand{\arraystretch}{1.2}
		\centering
		\begin{tabular}{clc}
			\hline
			No. & $\{k_1,k_2,\ldots,k_t\}$  & de Bruijn sequences based on  Equation (\ref{new3}) \\
			\hline
			$1$ & $\{1,7\}$         & $(0000001111110110100100110111010101100101000101111001110001100001)$ \\
			$2$ & $\{1, 2,7\}$      & $(0000001000011000101000111001001011001101001111010101110110111111)$ \\
			$3$ & $\{1, 2, 3,7\}$   & $(0000001000101001001101010111000111011000011001011011111100111101)$\\
			$4$ & $\{1,2,3,4,7\}$   & $(0000001001011011111100111101001100001101010001010111000111011001)$\\
			$5$ & $\{1,2,3,4,5,7\}$ & $(0000001010111000111011001001011011111100111101001100001101010001)$\\
			\hline
			No. & $\{k_1,k_2,\ldots,k_t\}$  &de Bruijn sequences based on  Equation (\ref{new4}) \\
			\hline
			$6$ & $\{1,7\}$         & $(0000001111110111100111000110110100110000101110101100101010001001)$ \\
			$7$ & $\{1, 3,7\}$      & $(0000001100111100101100011100001010100110111111011010111010001001)$ \\
			$8$ & $\{1, 3, 4,7\}$   & $(0000001100101101100011101111110101110011110000101010011010001001)$\\
			$9$ & $\{1,2,3,5,7\}$   & $(0000001001000101010011011010111010000110011110111111001011000111)$\\
			$10$ & $\{1,2,3,4,5,7\}$ & $(0000001001000101010011010000110010110110001110101110011110111111)$\\
			\hline
		\end{tabular}
	\end{table*}
	
	Taking the exhaustive approach incurs a steep penalty in memory requirement to store all qualified states in the corresponding cycles. Etzion and Lempel in \cite{Etzion84} stored many, not all, qualified states to perform cycle joining by successor rules. Their construction generates a large number, exponential in the order $n$, of de Bruijn sequences at the cost of raised memory demand.
	
	\subsection{Under the Shift Order}\label{subsec:shift}
	
	Imposing a {\it shift order} on the states in a given cycle yields a lot of feasible successor rules. We call a state whose first entry is $0$ a {\it leading zero state} or an LZ state in short. Analogously, a state whose last entry is $1$ is said to be an {\it ending one state} or an EO state.
	
	The necklace in a given cycle $(c_0,c_1,\ldots,c_{n-1}) \neq (1)$ must begin with $0$, \ie, its necklace is an LZ state. Here we define a \emph{special left shift operator}, denoted by
	$L_{\rm lz}$. Applied on a given LZ state $\mathbf{v}:=0,c_1,\ldots,c_{n-1}$ the operator $L_{\rm lz}$ outputs the first LZ state obtained by consecutive left shifts on $\mathbf{v}$. More formally,
	$L_{\rm lz} \, \mathbf{v}:=\mathbf{v}$ if $c_1,\ldots,c_{n-1}=1,\ldots,1$. Otherwise, let $1 \leq j < n$ be the least index such that $c_j=0$. Then
	\[
	L_{\rm lz} \, \mathbf{v} := 0, c_{j+1}, \ldots, c_{n-1}, 0, c_1, \ldots, c_{j-1}.
	\]
	
	Similarly, the necklace in any $C \neq (0)$ must end with $1$, \ie, it is an EO state. Given a state $\mathbf{u} := c_1,\ldots,c_{n-1},1$, the special operator $L_{\rm eo}$ fixes $\mathbf{u}$ if $c_1,\ldots,c_{n-1}:=0,\ldots,0$. Otherwise, let $1 \leq j < n$ be the least index such that $c_j=1$. Then
	\[
	L_{\rm eo} \, \mathbf{u} := c_{j+1},\ldots,c_{n-1},1,c_1,\ldots,c_{j-1},1.
	\]
	In other words, $L_{\rm eo}\, \mathbf{u}$ is the first EO state found upon consecutive left shifts on $\mathbf{u}$.
	
	For these two special operators, the convention is to let
	\begin{equation*}
		\begin{cases}
			L^0_{\rm lz} \, \mathbf{v} = \mathbf{v}, \\
			L^0_{\rm eo} \, \mathbf{u} = \mathbf{u},
		\end{cases}
		\mbox{ and} \quad
		\begin{cases}
			L^k_{\rm lz} \, \mathbf{v} =L_{\rm lz}^{k-1} (L_{\rm lz}\, \mathbf{v}),\\
			L^k_{\rm eo} \, \mathbf{u} = L_{\rm eo}^{k-1}(L_{\rm eo} \, \mathbf{u}),
		\end{cases}
		\mbox{ for } k > 0.
	\end{equation*}
	
	\begin{example}
		For $C=(001011)$ with $\mathbf{v} = 001011$, we have $L^1_{{\rm lz}} \, \mathbf{v} = 010110$ and $L^2_{{\rm lz}} \, \mathbf{v} = 011001$, whereas $L^1_{\rm eo} \, \mathbf{v} = 011001$ and $L^2_{\rm eo} \, \mathbf{v} = 100101$. \QEDB
	\end{example}
	
	Now we construct successor rules based on $L_{\rm lz}$ and $L_{\rm eo}$. 
	
	\begin{proposition}\label{prop:sf0}
		With arbitrarily chosen $2 \leq t \leq n$, we let $1=k_1< k_2<\cdots< k_{t}=n+1$ and $k_{t-1}<n$. For a state $\mathbf{c}:=c_0,c_1,\ldots,c_{n-1}$, let $\mathbf{v}:=0,c_1,\ldots,c_{n-1}$ and $\mathbf{u}:=c_1,\ldots,c_{n-1},1$.
		The following two successor rules generate de Bruijn sequences of order $n$.
		\begin{align}
			\rho_{\rm lz}(\mathbf{c}) & =
			\begin{cases}
				\overline{c_0} \mbox{, if } k_i \leq \wt(\overline{\mathbf{v}}) < k_{i+1} \mbox{ for some } i \\
				\quad \mbox{ and } \ L_{\rm lz}^{k_i-1} \, \mathbf{v} \mbox{ is a necklace}, \\
				c_0 \mbox{, otherwise}.
			\end{cases}\label{new3}\\
			\rho_{\rm eo}(\mathbf{c}) &=
			\begin{cases}
				\overline{c_0} \mbox{, if } k_i \leq \wt(\mathbf{u}) < k_{i+1} \mbox{ for some } i \\
				\quad \mbox{ and } \ L_{\rm eo}^{k_i-1} \, \mathbf{u} \mbox{ is a necklace}, \\
				c_0 \mbox{, otherwise}.
			\end{cases}\label{new4}
		\end{align}
	\end{proposition}
	
	In Proposition \ref{prop:sf0} we let $k_t = n+1$ for consistency since $\wt(\overline{\mathbf{v}})=n$ in $C=(0)$ and $\wt(\mathbf{u})=n$ in $C=(1)$. Each of these special cycles has only a single state. The reason to have $k_{t-1} < n$ is then clear. The correctness of Proposition~\ref{prop:sf0} comes from Theorem \ref{thm1} and the fact that the state satisfying the respective conditions in $\rho_{\rm lz}$ and $\rho_{\rm eo}$ is uniquely determined in the corresponding cycle. Examples of their output sequences are provided in Table~\ref{table:AtoE} for $n=6$.
	
	\begin{proposition}\label{prop:shift1314}
		Each of the successor rules $\rho_{\rm lz}$ in (\ref{new3}) and $\rho_{\rm eo}$ in (\ref{new4}) generates $2^{n-2}$ de Bruijn sequences of order $n$.
	\end{proposition}
	\begin{IEEEproof}
		We supply the proof for $\rho_{\rm lz}$ in (\ref{new3}), the other case being similar to argue. For each $1 \leq \ell < n$, there exists at least one cycle of the PCR of order $n$ having $\ell$ distinct LZ states. To verify existence, one can, \eg, inspect the cycle
		\[
		(\underbrace{00 \ldots 0}_{\ell} \underbrace{11 \ldots 1}_{n-\ell})\mbox{ for each }
		1 \leq \ell < n.
		\]
		On the other hand, taking all possible $2 \leq t\leq n$, there are $2^{n-2}$ distinct sets $\{1=k_1,  k_2,  \ldots,k_{t-1},k_t=n+1\}$ with $k_{t-1}<n$. Distinct sets provide distinct successor rules, producing $2^{n-2}$ inequivalent de Bruijn sequences in total.
	\end{IEEEproof}
	
	We are not quite done yet. Here are two more general successor rules whose validity can be routinely checked.
	\begin{proposition}\label{prop:ss}
		Let $k$ be a nonnegative integer. For a state $\mathbf{c}:=c_0,c_1,\ldots,c_{n-1}$, let $\mathbf{v}:=0,c_1,\ldots,c_{n-1}$ and $\mathbf{u}:=c_1,\ldots,c_{n-1},1$. The following successor rules generate de Bruijn sequences of order $n$.
		\begin{align}
			\rho(\mathbf{c}) &=
			\begin{cases}
				\overline{c_0} \mbox{, if } L_{\rm lz}^{k} \, \mathbf{v} \mbox{ is a necklace}, \\
				c_0 \mbox{, otherwise}.
			\end{cases}
			\label{new5}\\
			\rho(\mathbf{c}) &=
			\begin{cases}
				\overline{c_0} \mbox{, if } L_{\rm eo}^{k} \, \mathbf{u} \mbox{ is a necklace}, \\
				c_0 \mbox{, otherwise}.
			\end{cases}
			\label{new6}
		\end{align}
	\end{proposition}
	
	\begin{table*}
		\caption{A partial list of the resulting de Bruijn sequences from Proposition~\ref{prop:ss}, with $n=6$.}
		\label{table:EE}
		\renewcommand{\arraystretch}{1.2}
		\centering
		\begin{tabular}{ccc}
			\hline
			$k$  & de Bruijn sequences based on Equation (\ref{new5}) & Notes \\
			\hline
			0 & (0000001111110110100100110111010101100101000101111001110001100001) & PCR4 in~\cite{Gabric18}  \\
			1 & (0000001000011000101000111001001011001101001111010101110110111111)&
			{\tt granddaddy} \cite{Fred72} \\
			2 & (0000001000101001001101010111100111111011000011001011011100011101)& \\
			3 & (0000001001011101101001100001101111110011100011110101000101011001)& \\
			4 & (0000001011001101001001111110110111010101111000110000111001010001) & \\
			55 & (0000001101001111010100010101110110111111000111001001011001100001) & \\
			56 & (0000001000011101001001101010111100111111011001010001011011100011) & \\
			57 & (0000001000111101010110010010111011010011011111100111000011000101) & \\
			58 & (0000001001111110110111010101111000111001011000011001101000101001) & \\
			59 & (0000001010111000111011001001011011111100111101001100001101010001) & Equation (\ref{new1}) \\
			\hline
			$k$  & de Bruijn sequences based on Equation (\ref{new6}) & Notes  \\
			\hline
			0 & (0000001111110111100111000110110100110000101110101100101010001001) & PCR3(J1) in~\cite{Gabric18}\\
			1 & (0000001001111001101001011011001000111000101011111101110101000011) & Equation (\ref{new2})\\
			2 & (0000001100111100101100011100001010100110111111011010111010001001) & \\
			3 & (0000001001000101101100101010000111011111101011100111100011010011) & \\
			4 & (0000001100001010100011110111111001110001001101101001011101011001) & \\
			55 & (0000001001101001011011001000111010111001111110111100010101000011) & \\
			56 & (0000001100101111110111010110001111001110000101010011011010001001) & \\
			57 & (0000001001000101101111110110010101110101000011100011010011110011) & \\
			58 & (0000001100001010100011100010011011010111011111101001011001111001) & \\
			59 & (0000001001000101010011010000110010110110001110101110011110111111) & {\tt grandmama} \cite{Dragon18}\\
			\hline
		\end{tabular}
	\end{table*}
	
	\begin{proposition}\label{prop2}
		The number of distinct de Bruijn sequences of order $n$ produced by each of the rules in (\ref{new5}) and (\ref{new6}) is
		\begin{equation}
			\lcm(1,2,\ldots,n-1) \geq  (n-1)  \binom{n-2}{\left\lfloor \frac{n-2}{2} \right\rfloor} 
			\geq 2^{n-2}.
		\end{equation}
	\end{proposition}
	\begin{IEEEproof}
		We supply the counting for the successor rule in (\ref{new5}). We know from the proof of Proposition \ref{prop:shift1314} that, for each $1 \leq \ell < n$, there exists at least one cycle of the PCR of order $n$ having $\ell$ distinct LZ states.
		For a given $k$, we construct the system of congruences
		\begin{equation}\label{cong}
			\{k\equiv a_i \Mod{i} \mbox{ for } i \in \{1,2,\ldots,n-1\}\}.
		\end{equation}
		The number of resulting distinct de Bruijn sequences of order $n$ is equal to the number of solvable systems of congruences in (\ref{cong}). The sequences are distinct because different nonempty subsets of $\{a_1,\ldots,a_{n-1}\}$, whose corresponding systems are solvable, lead to different choices for the uniquely determined states in the respective cycles. By Generalized Chinese Remainder Theorem, the number is $\lcm(1,2,\ldots,n-1)$. From~\cite[Section 2]{Farhi} we get the lower bounds
		\[
		(n-1)  \binom{n-2}{\left\lfloor \frac{n-2}{2} \right\rfloor} 
		\geq 2^{n-2}.
		\]
	\end{IEEEproof}
	
	Proposition \ref{prop:ss} includes the constructions of de Bruijn sequences from the PCR of order $n$ in~\cite{Gabric18} as special cases. Taking $k \in \{0,1,\lcm(1,2,\ldots,n-1)-1\}$ in (\ref{new5}), for instance, outputs three sequences, namely the PCR$4$ in \cite{Gabric18}, {\tt granddaddy}, and a sequence from (\ref{new1}), respectively. Using (\ref{new6}) with $k \in \{0,1,\lcm(1,2,\ldots,n-1)-1\}$ yields the sequence PCR$3$\,(J$1$) in \cite{Gabric18}, a sequence from (\ref{new2}), and {\tt grandmama}, respectively.
	
	\begin{example}
		When $n=6$, each successor rule in Proposition~\ref{prop:ss} yields $60$ distinct de Bruijn sequences. Table~\ref{table:EE} lists only $10$ of the $60$. We note their connection to known sequences to illustrate the generality of our approach. \QEDB
	\end{example}
	
	For the successor rules in Propositions \ref{prop:sf0} and \ref{prop:ss}, generating the next bit means checking if a state is a cycle's necklace by repeated simple left shifts. This can be done in $O(n)$ time and $O(n)$ space.
	
	We generalize Proposition \ref{prop:ss} to define more successor rules.
	\begin{theorem}\label{thm:gg}
		Let $g(k)$ be an arithmetic function
		\begin{equation}\label{eq:gk}
			g(k): \{1,2,\ldots,n\} \mapsto \{0,1,\ldots,k-1\}.
		\end{equation}
		As before, for any $\mathbf{c}:=c_0,c_1,\ldots,c_{n-1}$, let $\mathbf{v}:=0,c_1,\ldots,c_{n-1}$ and $\mathbf{u} := c_1,\ldots,c_{n-1},1$. The following successor rules generate de Bruijn sequences of order $n$.
		\begin{align}
			\rho^g_{{\rm lz}}(\mathbf{c}) &=
			\begin{cases}
				\overline{c_0} \mbox{, if } L_{\rm lz}^{g(\wt(\overline{\mathbf{v}}))} \, \mathbf{v} \mbox{ is a necklace}, \\
				c_0 \mbox{, otherwise}.
			\end{cases}
			\label{new7}\\
			\rho^g_{{\rm eo}}(\mathbf{c}) &=
			\begin{cases}
				\overline{c_0} \mbox{, if } L_{\rm eo}^{g(\wt(\mathbf{u}))} \, \mathbf{u} \mbox{ is a necklace}, \\
				c_0 \mbox{, otherwise}.
			\end{cases}\label{new8}
		\end{align}
	\end{theorem}
	
	For a cycle with $1\leq \ell\leq n-1$ distinct LZ states, there are $\ell$ distinct ways to choose the uniquely determined state according to $g(\ell)$. The counting for $\ell$ distinct EO states is identical. It is then straightforward to confirm that each successor rule in Theorem \ref{thm:gg} can generate $(n-1)!$ distinct de Bruijn sequences of order $n$ by using all possible $g(\ell)$.
	
	\subsection{The Feedback Functions of the Resulting Sequences}
	Let $\mathbf{x}:=x_0,x_1,\ldots,x_{n-1}$ be any state. We briefly discuss the feedback functions of the de Bruijn sequences produced earlier in this section. Their form is
	\[
	f(\mathbf{x})=
	\begin{cases}
		\overline{x_0} \mbox{, if } \mathbf{x} \mbox{ satisfies the specified condition,} \\
		x_0 \mbox{, otherwise.}
	\end{cases}
	\]
	Let $E$ be the set of states ${\bf v}=v_0,v_1,\ldots,v_{n-1}$ such that each conjugate pair $({\bf v},\widehat{{\bf v}})$ is used in generating the corresponding de Bruijn sequence. The feedback function of the resulting sequence is therefore
	\begin{align}\label{eq:FF1}
		f(x_0,x_1,\ldots,x_{n-1}) & = x_0+h(x_1,\ldots,x_{n-1}) \mbox{, with} \notag \\
		h(x_1,\ldots,x_{n-1}) & =
		\sum_{\mathbf{v}\in E} 
		\left( \prod_{1 \leq i < n} (x_i + \overline{v_i}) \right).
	\end{align}
	Hence, determining $f$ requires computing $h$ such that
	\[
	h(x_1,\ldots,x_{n-1})=
	\begin{cases}
		1 \mbox{, if }  x_1,\ldots,x_{n-1} \mbox{ meets the condition,}\\
		0 \mbox{, otherwise.}
	\end{cases}
	\]
	Since the resulting de Bruijn sequences come from joining all of the cycles in $\Omega(f_{\rm PCR})$, we have $\wt(h) = Z_n-1$.
	
	The following proposition will soon be useful.
	\begin{proposition}\label{prop13}
		The feedback function of the successor rule
		\begin{equation}\label{eq:FF2}
			\rho(\mathbf{x})=
			\begin{cases}
				1 \mbox{, if } \mathbf{x} \mbox{ is a necklace},\\
				0 \mbox{, otherwise}
			\end{cases} 
			\mbox{ is } 
			f_{\rho}(\mathbf{x}) := \prod_{i=1}^{n-1} f_i(\mathbf{x}),
		\end{equation}
		where
		\begin{equation*}
			f_i(\mathbf{x}) =\overline{x_0} \cdot x_i + \overline{(x_0 + x_i)} \cdot \overline{x_1} \cdot x_{i+1} + \ldots 
			+ \overline{(x_0+x_i)} \cdots \overline{(x_{n-2} + x_{i+n-2})} \cdot \overline{x_{n-1}} \cdot x_{n-1+i} 
			+\overline{(x_0 + x_i)} \cdots \overline{(x_{n-1} + x_{n-1+i})}.
		\end{equation*}
	\end{proposition}
	
	\begin{IEEEproof}
		The state $\mathbf{x}:=x_0,x_1,\ldots,x_{n-1}$ is a necklace if and only if it is lexicographically least in the set of all of its shifts. Let $\mathbf{x}_{i}:=x_i,x_{i+1},\ldots,x_{i+n-1}$, where $1 \leq i < n$ and the subscripts are computed modulo $n$. The notation $\wedge$ stands for the logical AND. Then $\mathbf{x} \preceq_{\rm lex} \mathbf{x}_i$ if and only if one and only one of the following conditions holds.
		\begin{align*}
			0 =x_0 < x_i=1  &\iff \overline{x_0} \cdot x_i=1, \\
			(x_0 = x_i) \wedge (x_1 < x_{i+1}) &\iff
			\overline{(x_0+x_i)} \cdot \overline{x_1} \cdot x_{i+1}=1, \\
			\cdots \quad \cdots \quad & \iff \quad \cdots \quad \cdots, \\
			(x_0 = x_i) \wedge \ldots \wedge (x_{n-2} = 
			x_{i + n -2}) \wedge (x_{n-1} < x_{i+n-1}) & \iff \overline{(x_0 + x_i)} \cdots 
			\overline{(x_{n-2} + x_{i+n-2})} \cdot \overline{x_{n-1}} \cdot x_{i+n-1}=1, \\
			(x_0 = x_i) \wedge \ldots \wedge (x_{n-2} = 
			x_{i + n -2}) \wedge (x_{n-1} = x_{i+n-1}) &\iff
			\overline{(x_0 + x_i)} \cdots \overline{(x_{n-1} + x_{i+n-1})}=1.
		\end{align*}
		Hence, $\mathbf{x} \preceq_{\rm lex} \mathbf{x}_i$ if and only if $f_i=1$ for all $1\leq i < n$.
	\end{IEEEproof}
	
	The next two corollaries to Proposition~\ref{prop13} give the respective feedback functions of the stated successor rules.
	\begin{corollary}\label{cor1}
		Let $\mathbf{c}:=c_0,c_1,\ldots,c_{n-1}$. The feedback function of the de Bruijn sequence built by the successor rule
		\[
		\rho(\mathbf{c})=
		\begin{cases}
			\overline{c_0} \mbox{, if } 0,c_1,\ldots,c_{n-1} \mbox{ is a necklace},\\
			c_0 \mbox{, otherwise,}
		\end{cases}
		\mbox{ is }
		f(x_0,x_1,\ldots,x_{n-1}) = x_0+ f_{\rho}(0,x_1,\ldots,x_{n-1}).
		\]
	\end{corollary}
	
	\begin{corollary}\label{cor2}
		Let $\mathbf{c}:=c_0,c_1,\ldots,c_{n-1}$. The feedback function of the successor rule
		\[
		\rho(\mathbf{c})=
		\begin{cases}
			\overline{c_0} \mbox{, if } c_1,\ldots,c_{n-1},1 \mbox{ is a necklace},\\
			c_0 \mbox{, otherwise,}
		\end{cases}
		\mbox{ is }
		f(x_0,x_1,\ldots,x_{n-1})=x_0 + f_{\rho}(x_1,\ldots,x_{n-1},1).
		\]
	\end{corollary}
	
	The feedback function of the resulting de Bruijn sequence built by the other successor rules that we have discussed above can be inferred from a similar analysis on the corresponding Boolean logical operations. The details are omitted here.
	
	\section{Successor Rules from Pure Summing Registers}\label{sec:PSR}
	
	This section studies how to generate de Bruijn sequences by applying the CJM on the PSR of any order $n$. The strategy is to define several distinct relations or orders on the cycles before deploying them in constructing new successor rules. Let $B$ be a statement which guarantees that the resulting sequence is de Bruijn, given that the FSR is the PSR of order $n$. Hence, analogous to $\rho_{A}$ in (\ref{equ:1}), for any state $\mathbf{c}:=c_0,c_1,\ldots,c_{n-1}$, we define the successor rule $\rho_{B}$ by
	\begin{equation}\label{equ:PSR}
		\rho_{B}(\mathbf{c})=
		\begin{cases}
			1+ \sum_{i=0}^{n-1} c_i \mbox{, if } \mathbf{c} \mbox{ satisfies } B,\\
			\sum_{i=0}^{n-1} c_i \mbox{, otherwise.}
		\end{cases}
	\end{equation}
	
	\subsection{The Run Order on the Pure Summing Register}\label{subsec:order}
	For a binary periodic sequence, a run of $k$ consecutive $0$s preceded and followed by a $1$ is {\it a run of $0$s of length $k$}. A run of $1$s of length $k$ is defined analogously. The convention is to fix $k =\infty$ and $k=0$ as the respective lengths of runs of $0$s in $(0)$ and in $(1)$. This subsection imposes a new order on the cycles of the PSR based on their runs of $0$s.
	
	For a given $n$, let $r$ be the number of cycles in $\Omega_{f_{\rm PSR}}$. Given cycles $C_i \neq C_j$ with $1 \leq i,j \leq r$, we say that $C_j \prec_{\rm rz}C_i$ in the {\it run order} if and only if the maximal length of runs of $0$s in $C_j$ is less than the maximal length of runs of $0$s in $C_i$. The subscript ${\rm rz}$ signifies that the arrangement is based on the {\it run of zeros}. Strictly speaking, the run order is just a transitive relation since it is not necessary to define how $C_i$ is related to itself.
	
	\begin{theorem}\label{thm:no}
		The following can serve as $B$ to define $\rho_{B}$ in Equation (\ref{equ:PSR}).
		\begin{quote}
			The $(n+1)$-stage state $c_1,\ldots,c_{n-1},1+\sum_{i=1}^{n-1} c_i,1$ is uniquely determined in the corresponding cycle that begins with a maximal length run of $0$s.
		\end{quote}
	\end{theorem}
	
	\begin{IEEEproof}
		Recall that the length of run of $0$s in $(0)$ is $\infty$. By Theorem~\ref{thm:order}, it suffices to show that for each $C_i \neq (0)$, there is a uniquely determined state whose conjugate state is in $C_j$, with $C_j \succ_{\rm rz} C_i$.
		Any nonzero cycle
		\[
		C_i := \left(1,c_1,\ldots,c_{n-1},1+\sum_{i=1}^{n-1} c_i\right)
		\]
		has at least one state with the property that a maximal length run of $0$s starts at $c_1$. Suppose that $1,c_1,\ldots,c_{n-1}$ has been uniquely identified. Then its conjugate, namely $0,c_1,\ldots,c_{n-1}$, must be in
		\[
		C_j := \left(0,c_1,\ldots,c_{n-1},\sum_{i=1}^{n-1} c_i\right)
		\]
		with a larger maximal length run of $0$s. Thus, $C_j \succ_{\rm rz} C_i$.
	\end{IEEEproof}
	
	\begin{remark}
		The run order here is well-defined for arbitrary FSRs. We can use it to generate de Bruijn sequences by joining the cycles of an arbitrary FSR based on a similarly defined successor rules. Efficiency is another matter altogether since the cycle structure may be harder to manage if the choice of the FSR is not done judiciously. \QEDB
	\end{remark}
	
	How to efficiently determine a unique state in a nonzero $C \in \Omega_{f_{\rm PSR}}$ of order $n$? An $(n+1)$-stage state that starts with a run of $0$s with maximal length must exist because the necklace satisfies this condition. If $C = (1)$, which happens whenever $n$ is odd,  then the maximal length of a run of $0$s is understood to be $0$ and we use $1,\ldots,1$ as the required $(n+1)$-stage. Suppose that $\mathbf{v}$ is an $(n+1)$-stage state that starts with a maximal length run of $0$s. A new operator $L_{\rm rz}$ on $\mathbf{v}$ is defined such that $L_{\rm rz} \, \mathbf{v}$ is the next state that also starts with a maximal length run of $0$s obtained by repeated applications of the left shift $L$ on $\mathbf{v}$. Let $L_{\rm rz}^k \, \mathbf{v} = L^{k-1}_{\rm rz}(L_{\rm rz} \, \mathbf{v})$ for any positive integer $k$.
	
	We can now propose several distinct successor rules simply by specifying how to uniquely determine the $(n+1)$-stage state in Theorem \ref{thm:no}.
	
	\begin{proposition}\label{prop:psrneck}
		Let $k$ be a nonnegative integer. For a given state $\mathbf{c}:=c_0,c_1,\ldots,c_{n-1}$, let
		\[
		\mathbf{v} :=c_1,\ldots,c_{n-1},1+\sum_{i=1}^{n-1} c_i,1.
		\]
		We call $\mathbf{v}$ a \emph{nice state} if it starts with a maximal length run of $0$s and $L^k_{\rm rz} \, \mathbf{v}$ is a necklace. Then the successor rule
		\begin{equation}\label{new9}
			\rho_{\rm nice}(\mathbf{c}) =
			\begin{cases}
				1+ \sum_{i=0}^{n-1} c_i \mbox{, if } \mathbf{v} \mbox{ is nice},\\
				\sum_{i=0}^{n-1} c_i \mbox{, otherwise},
			\end{cases}
		\end{equation}
		generates de Bruijn sequences of order $n$.
	\end{proposition}
	
	If $k=0$ in Proposition \ref{prop:psrneck}, then the rule simplifies to
	\begin{equation}\label{equ:psrneck1}
		\rho_{\rm nice}(\mathbf{c})=
		\begin{cases}
			1+ \sum_{i=0}^{n-1} c_i \mbox{, if } \mathbf{v} \ \mbox{is a necklace},\\
			\sum_{i=0}^{n-1} c_i \mbox{, otherwise.}
		\end{cases}
	\end{equation}
	
\begin{figure}
		\centering
		\begin{tikzpicture}
			[
			> = stealth,
			shorten > = 3pt,
			auto,
			node distance = 2.2cm,
			semithick
			]
			
			\tikzstyle{every state}=
			\node[rectangle,fill=white,draw,rounded corners,minimum size = 4mm]
			
			\node[state] (1) {$C_1$};
			\node[state] (2) [right of=1] {$C_2$};
			
			\node[state] (3) [below of=1] {$C_6$};
			\node[state] (4) [left of=3]{$C_3$};
			\node[state] (5) [right of=3] {$C_8$};
			
			\node[state] (6) [right of=5] {$C_5$};
			\node[state] (7) [below of=4] {$C_4$};
			\node[state] (8) [right of=7] {$C_9$};
			\node[state] (9) [below of=6] {$C_7$};
			
			\node[state] (10) [right of=8] {$C_{10}$};
			
			\path[->] (1) edge node [right] {$101111$} (3);
			\path[->] (4) edge node [above] {$100111$} (3);
			
			\path[->] (2) edge node [right] {$101010$} (5);
			\path[->] (6) edge node [above] {$100101$} (5);
			\path[->] (9) edge node [right] {$100010$} (5);
			
			\path[->] (3) edge node [left] {$100011$} (8);
			\path[->] (5) edge node [right] {$100001$} (8);
			\path[->] (7) edge node [below] {$100110$} (8);
			
			\path[->] (8) edge node [below] {$100000$} (10);
			
		\end{tikzpicture}
		\caption{The spanning tree produced by the $\rho_{\rm nice}$ in Proposition \ref{prop:psrneck} when applied to the PSR of order $6$ with $c_0,c_1,\ldots, c_5 = 101010$ chosen in $C_2$. The edge label from $C_i$ to $C_j$ is the chosen $c_0,c_1,\ldots,c_5$ in $C_i$ whose conjugate is in $C_j$.}
		\label{fig:exPSR}
	\end{figure}
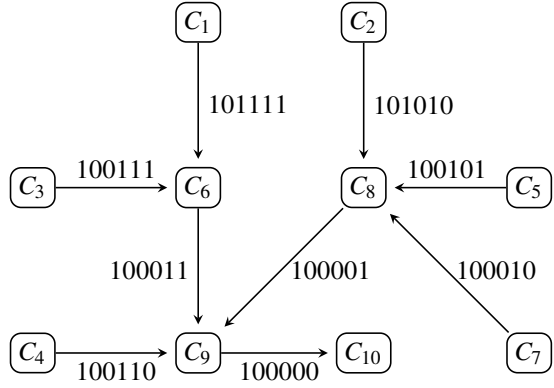

	\begin{example}\label{ex:nk}
		We label the $10$ cycles generated by the PSR of order $6$ as
		\begin{align*}
			& C_1=(0111111), ~C_2=(0101011), ~C_3=(0011101), ~C_4=(0011011), ~C_5=(0010111),\\
			& C_6=(0001111),~ C_7=(0001001), ~C_8=(0000101), ~C_9=(0000011),~ C_{10}=(0000000).
		\end{align*}
		All, except for $C_{10}$, have least period $7$. In the run order $C_{10}$ is the largest. The maximal lengths of runs of $0$s in $C_8$ and $C_4$ are $4$ and $2$ respectively, implying $C_4 \prec_{\rm rz} C_8$.
		
		The maximal length of run of $0$s in $C_2$ is $1$ with $3$ distinct $7$-stage states that start with $0$, namely $\mathbf{v}=0101011$, $L_{\rm rz} \, \mathbf{v}=0101101$, and $L^2_{\rm rz} \mathbf{v} =0110101$, since $L^3_{\rm rz}\,\mathbf{v} = \mathbf{v}$. Each of these $3$ states can be chosen to be the uniquely determined state. Each $C_i$, for $i \notin \{2,10\}$, has only one choice for a state that starts with the respective longest run of $0$s. Proposition \ref{prop:psrneck} yields three distinct de Bruijn sequences which are presented in Table \ref{table:FandG}. The spanning tree when $\mathbf{v}$ is chosen, \ie, when $c_0,c_1,\ldots,c_5 = 101010$ is given in Figure~\ref{fig:exPSR}. \QEDB
	\end{example}
	
	\subsection{The Necklace Order on the Pure Summing Register}\label{subsec:neck}
	
	This subsection presents another general method to construct successor rules which can generate de Bruijn sequences based on the PSR of order $n$. Given $n$, each cycle in
	$\Omega_{f_{\rm PSR}}$ is $(n+1)$-periodic and, hence, can be written as $(c_0,c_1,\ldots,c_n)$. We define a new total order, which we name the {\it necklace order} denoted by $\prec_{\rm nk}$, on the cycles. Given $C_i \neq C_j$, we say $C_i \prec_{\rm nk} C_j$ if and only if the necklace of $C_i$ is lexicographically less than that of $C_j$.
	
	The companion state of $c_0,\ldots,c_{n-1}$ in $C_i =(c_0,c_1,\ldots,c_{n})$ is in $C_j =(c_0,\ldots,c_{n-2},\overline{c_{n-1}},\overline{c_n})$, making $C_i$ and $C_j$ adjacent. Our task is to determine a state in $C_i$ whose companion state is in $C_j$ with $C_j \prec_{\rm nk} C_i$.
	
	\begin{lemma}\label{lem:1}
		For a given $n$, let $C_i:=(c_0,c_1,\ldots,c_{n-1},1)$ and $C_j: = (\overline{c_0},c_1,\ldots,c_{n-1},0)$ be adjacent nonzero cycles in $\Omega_{f_{\rm PSR}}$.
		\begin{enumerate}
			\item If $c_0,c_1,\ldots,c_{n-1},1$ is the unique EO state, that is, it is the necklace in $C_i$, then $C_j \prec_{\rm nk} C_i$.
			\item If there are two or more EO states in $C_i$ and the necklace in $C_i$ is not $c_0,c_1,\ldots,c_{n-1},1$, then $C_j \prec_{\rm nk} C_i$.
		\end{enumerate}
	\end{lemma}
	
	\begin{IEEEproof}
		In the first case, $C_i$ is either $(1)$ or has the form
		\[
		(\underbrace{0,\ldots,0}_t, 1, \, \ldots \, ,  \underbrace{0,\ldots,0}_t, 1).
		\]
		If $C_i =(1)$, then $C_j=(0,1,\ldots,1,0)$, making $C_j \prec_{\rm nk} C_i$. If $C_i \neq (1)$, then it is easy to confirm that the maximal length of the run of $0$s in $C_j$ is larger than that of $C_i$. Thus, $C_j \prec_{\rm nk} C_i$ as well.
		
		For the second case, suppose that the necklace of $C_i$ is of the form
		\[
		c_i,\ldots,c_{n-1},1,c_0,\ldots,c_{i-1}
		\]
		for some positive integer $i$. Then there exists a state in $C_j$ with the form
		\[
		c_i,\ldots,c_{n-1},0,\overline{c_0},\ldots,c_{i-1}
		\]
		which is lexicographically less than the necklace of $C_i$. Thus, $C_j \prec_{\rm nk} C_i$.
	\end{IEEEproof}
	
	In a similar manner we can prove the following lemma, which will be used in the proof of Theorem~\ref{thm2}.
	\begin{lemma}\label{lem:15}
		For a given $n$, let $C_i:=(0,c_1,\ldots,c_{n-1},1)$, with $c_1,\ldots,c_{n-1} \neq 1,\ldots,1$, and $C_j: = (1,c_1,\ldots,c_{n-1},0)$ be adjacent nonzero cycles in $\Omega_{f_{\rm PSR}}$.
		\begin{enumerate}
			\item If $0,c_1,\ldots,c_{n-1},1$ is the unique LZ state, that is, it is the necklace in $C_i$, then $C_j \prec_{\rm nk} C_i$.
			\item If there are two or more LZ states in $C_i$ and the necklace in $C_i$ is not $0,c_1,\ldots,c_{n-1},1$, then $C_j \prec_{\rm nk} C_i$.
		\end{enumerate}
	\end{lemma}
	
	Combining Theorem \ref{thm:order} and Lemma \ref{lem:1} results in the next theorem.
	
	\begin{theorem}\label{psr:nk}
		Let $\mathbf{c}:=c_0,c_1,\ldots,c_{n-1}$ and $\mathbf{v}:=1 + \sum_{i=1}^{n-1},c_1,\ldots,c_{n-1},1$. The successor rule $\rho_{B}$ defined as
		\begin{equation}
			\rho_{B}(\mathbf{c})=
			\begin{cases}
				1+ \sum_{i=0}^{n-1} c_i \mbox{, if } \mathbf{v} \mbox{ satisfies  Condition}\ B,\\
				\sum_{i=0}^{n-1} c_i \mbox{, otherwise,}
			\end{cases}
		\end{equation}
		generates a de Bruijn sequence. Condition $B$ consists of any of the following two requirements.
		\begin{enumerate}
			\item The state $\mathbf{v}$ is the only EO state in its cycle.
			\item Among the two or more EO states in the cycle, $\mathbf{v}$ is not the necklace, but it can be uniquely identified.
		\end{enumerate}
	\end{theorem}
	
	Theorem \ref{psr:nk} allows us to generate $3^5 \cdot 5 = 1, \, 215$ distinct de Bruijn sequences from the PSR of order $n=6$. Some of the successor rules, however, are rather ad hoc and not easy to generate. The next proposition gives a simple rule to efficiently and uniquely identify an EO state which is not a necklace by using the operator $L_{\rm eo}$ defined earlier.
	
	\begin{proposition}\label{prop:nk}
		Let $k$ be a positive integer. Given a state $\mathbf{c}:=c_0,c_1,\ldots,c_{n-1}$, let $\mathbf{v} := 1+\sum_{s=1}^{n-1}c_s,c_1,\ldots,c_{n-1},1$. The following rule generates a de Bruijn sequence of order $n$.
		\begin{equation}\label{new11}
			\rho_{k}(\mathbf{c}) =
			\begin{cases}
				1+\sum_{i=0}^{n-1}c_i \mbox{, if } L^k_{\rm eo} \, \mathbf{v} \neq \mathbf{v} \mbox{ and } 
				L^k_{\rm eo} \, \mathbf{v} \mbox{ is a necklace},\\
				1+\sum_{i=0}^{n-1}c_i \mbox{, if } L^{k}_{\rm eo}\, \mathbf{v} = \mathbf{v} \mbox{ and } L_{\rm eo} \, \mathbf{v} \mbox{ is a necklace}, \\
				\sum_{i=0}^{n-1}c_i \mbox{, otherwise}.
			\end{cases}
		\end{equation}
	\end{proposition}
	
	\begin{IEEEproof}
		Assume  $L^{k}_{\rm eo}\,\mathbf{v} =\mathbf{v}$ and $L_{\rm eo} \, \mathbf{v}$ is the necklace in $C$. If a cycle $C$ contains only one EO state, then $\mathbf{v}$ must be this state and it is the necklace, which is uniquely determined.   If $C$ has $\ell > 1$ distinct EO states, then we note that $L^{k}_{\rm eo} \, \mathbf{v} = \mathbf{v}$ if and only if $\ell \mid k$. If $L^{k}_{\rm eo} \, \mathbf{v} = \mathbf{v}$ and $L_{\rm eo} \, \mathbf{v}$ is the necklace, then $\mathbf{v}$ is not the necklace, but $\mathbf{v}$ is uniquely determined. If $L^k_{\rm eo} \, \mathbf{v} \neq \mathbf{v}$ and $L^k_{\rm eo} \, \mathbf{v}$ is the necklace, then $\mathbf{v}$ is uniquely determined, although it cannot be the necklace. Thus, by Theorem \ref{psr:nk}, the rule $\rho_{k}$ in (\ref{new11}) generates a de Bruijn sequence.
	\end{IEEEproof}
	
	We enumerate the number of distinct de Bruijn sequences from $\rho_k$ in (\ref{new11}) based on the number of valid values that $k$ can take.
	
	\begin{proposition}\label{prop:18}
		The number of de Bruijn sequences of order $n \geq 2$ that can be generated from Proposition \ref{prop:nk} is
		\begin{equation}\label{eq:Number}
			\lcm(2,4,\ldots,2\lfloor{n}/{2}\rfloor)-1\geq 2^{\lfloor{n}/{2}\rfloor}-1.
		\end{equation}
	\end{proposition}
	
	\begin{IEEEproof}
		Since the proposition clearly holds for $n \in \{2,3\}$, we treat $n \geq 4$.
		
		Let the numbers of distinct EO states in the cycles generated by the PSR of a fixed order $n \geq 2$ 
		be listed as $b_1,b_2,\ldots,b_t$, with $b_j > 1$ for $1 \leq j \leq t$. We exclude $b_j \in \{0,1\}$ for obvious reason. By the properties of the PSR, for each $1 \leq i \leq \lfloor{n}/{2}\rfloor$, there exists some $1 \leq s \leq t$ such that $2 i = b_s$. Since $\wt(C_{\rm PSR})$ is even, if $b_j > 1$ is odd, for some $1 \leq j \leq t$, then $b_j < \lfloor{n}/{2}\rfloor$ and $b_j \mid 2 i$, for some $i$. Hence,
		\[
		\lcm(b_1,b_2,\ldots,b_t)=\lcm(2,4,\ldots,2\lfloor{n}/{2}\rfloor).
		\]
		
		For a given $k$, we construct a vector of integers
		\begin{equation}\label{equ:nk11}
			\left(a_1',a_2',\ldots,a_t'\right) \mbox{ with } a_j' \equiv k \Mod{b_j} \mbox{ for all } 1 \leq j \leq t.
		\end{equation}
		By Theorem~\ref{thm:gcrt}, the corresponding vectors $\left(a_1',a_2',\ldots,a_t'\right)$ in (\ref{equ:nk11}) are distinct for distinct choices of $k$, as $k$ ranges from $1$ to $\lcm(b_1,b_2,\ldots,b_t)$. Thus, the number of such vectors is $\lcm(2, 4, \ldots, 2\lfloor{n}/{2}\rfloor)$.
		
		From each vector $\left(a_1',a_2',\ldots,a_t'\right)$, we construct a new vector 
		\begin{equation}\label{equ:nk}
			\left(a_1,a_2,\ldots,a_t\right) 
			\mbox{ with }
			a_{j}=
			\begin{cases}
				a_j' \mbox{, if } a_j' \neq 0, \\
				1 \mbox{, if } a_j' = 0,
			\end{cases}
			\mbox{ for all } 1 \leq j \leq t.
		\end{equation}
		Following the rule $\rho_{k}$ in (\ref{new11}), the uniquely determined state $\mathbf{v}$ in the cycle $C_{\rm PSR}$ with $b_j$ distinct EO states satisfies
		\begin{quote}
			$L^{a_{j}}_{\rm eo} \, \mathbf{v}$ is the necklace in $C_{\rm PSR}$.
		\end{quote}
		Thus, distinct vectors $\left(a_1,a_2,\ldots,a_t\right)$ in (\ref{equ:nk}) yield distinct de Bruijn sequences. We now enumerate the number of distinct vectors $\left(a_1,a_2,\ldots,a_t\right)$ in (\ref{equ:nk}). 
		
		If $k=1$ or $k = \lcm(b_1,b_2,\ldots,b_t)= \lcm(2,4,\ldots,2\lfloor{n}/{2}\rfloor)$, then $\left(a_1,a_2,\ldots,a_t\right)= (1,1,\ldots,1)$. Conversely, we claim that the only $k$, with $1 \leq k < \lcm(2,4,\ldots,2\lfloor{n}/{2}\rfloor)$, that satisfies $\left(a_1,a_2,\ldots,a_t\right)= (1,1,\ldots,1)$ is $k=1$. The condition implies that $a_j' \equiv k \Mod{b_j} \in \{0,1\}$, for $1\leq j\leq t$. By Theorem~\ref{thm:gcrt}, if $b_j > 1$ is odd, then $k \Mod{b_j} = k\Mod{2b_j}$ because $b_j \mid (k\Mod{b_j}-k\Mod{2b_j})$. Similarly, $k\Mod{2j} = k\Mod{2i}$, because $2 \mid (k\Mod{2j}-k\Mod{2i})$. Having
		\[
		k\Mod{b_1}=\ldots=k\Mod{b_t}=1 \mbox{ forces } k=1.
		\]
		
		Let $1 < k < \lcm(2,4,\ldots,2\lfloor{n}/{2}\rfloor)$ and $1 \leq \ell \leq t$. Since $k \neq 1$, there exists some $\ell$ such that $a_{\ell}=a_{\ell}'= k \Mod{b_{\ell}} > 1$. If $b_{\ell}$ is odd, then it is immediate to confirm that $k \Mod{2b_{\ell}} > 1$. Without loss of generality, we can assume that $b_{\ell}$ is even. We construct a bijection between $\left\{\left(a_1',a_2',\ldots,a_t'\right)\right\}$ and $\left\{\left(a_1,a_2,\ldots,a_t \right)\right\}$ by showing that each possible vector $\left(a_1,a_2,\ldots,a_t\right)\neq(1,1,\ldots,1)$ uniquely determines a vector $\left(a_1',a_2',\ldots,a_t'\right)$. If $a_j > 1$ for all $j$, then we are done. Otherwise, if $a_j = 1$, then $a_j' \in \{0,1\}$. In this case, if $b_j$ is odd, then we consider $b_s :=2 b_j$ and $a_s$. If $a_s' = a_s >1$, then $a_j'$ is fixed because $b_j \mid (a_s'-a_j')$. If $a_s =1$, then, since $b_{\ell}$ is even and $a_{\ell} = a_{\ell}' = k\Mod{b_{\ell}} >1$, we can determine $a_s'$ from the fact that $2 \mid (a_{\ell}'-a_s')$. Once this is done, $a_j'$ is accordingly identified.
		
		The results that we have established imply that, for $1\leq k < \lcm(2,4,\ldots,2\lfloor{n}/{2}\rfloor)$, distinct choices for $k$ yield distinct de Bruijn sequences upon applying the rule $\rho_{k}$ in (\ref{new11}).
		Thus, the number of distinct de Bruijn sequences is
		\begin{equation}
			\lcm \left(2,4, \ldots, 2 \lfloor{n}/{2}\rfloor\right)-1
			= 2 \cdot \lcm \left(1, 2, \ldots,\lfloor{n}/{2}\rfloor\right) -1
			\geq 2^{\lfloor{n}/{2}\rfloor}-1.
		\end{equation}
		The lower bound comes from a result in \cite[Section 2]{Farhi}.
	\end{IEEEproof}
	
	One can slightly modify the condition in Proposition \ref{prop:nk} to generate even more de Bruijn sequences. When $L^{k}_{\rm eo} \, \mathbf{v}=\mathbf{v}$, there are some quite obvious ways to determine the unique state in the corresponding cycle. The details can be worked out routinely and, hence, are omitted here.
	
	Another approach is to simplify $\rho_k$ by restricting the choices for $k$. Given $\mathbf{c}:=c_0,c_1,\ldots,c_{n-1}$, one constructs a new $(n+1)$-stage state $d_0,d_1,\ldots,d_{n-1},1$ such that 
	\[
	d_0 = 1+\sum_{i=1}^{n-1} c_i \mbox{ and } 
	d_i=c_i \mbox{ for } i \in \{1,\ldots,n-1\}.
	\]
	Let $0 \leq s, t < n$ be, respectively, the largest and smallest indices such that $d_s = d_t = 1$. Such indices exist since the state contains at least two $1$s. Let 
	\begin{align*}
		\mathbf{v}_s &:= d_{s+1},\ldots,d_{n-1},1,d_0,\ldots,d_{s} \mbox{ and} \\
		\mathbf{v}_t &:= d_{t+1},\ldots,d_{n-1},1,d_0,\ldots,d_{t}.    
	\end{align*}
	We then define the successor rule
	\begin{equation}\label{equ:nk1}
		\rho_{s}(\mathbf{c})=
		\begin{cases}
			1+ \sum_{i=0}^{n-1} c_i \mbox{, if } \mathbf{v}_s  \mbox{ is a necklace},\\
			\sum_{i=0}^{n-1} c_i \mbox{, otherwise.}
		\end{cases}
	\end{equation}
	The rule $\rho_{t}$ is exactly the same with $\mathbf{v}_t$ replacing $\mathbf{v}_s$.
	
	\begin{table*}
		\caption{Distinct de Bruijn sequences from Propositions~\ref{prop:psrneck}, \ref{prop:nk} and \ref{prop:mix}, with $n=6$.}
		\label{table:FandG}
		\renewcommand{\arraystretch}{1.2}
		\centering
		\begin{tabular}{ccc}
			\hline
			$k$  & de Bruijn sequences based on Equation (\ref{new9}) & Notes \\
			\hline
			0 & (0000001001000101110010101101010000111010011111101111000110110011) & Also from Equation (\ref{equ:psrneck1}) \\
			1 & (0000001001000101110010100001110100111111011110001101010110110011) &  \\
			2 & (0000001001000101101010111001010000111010011111101111000110110011) &  \\
			\hline
			$k$  & de Bruijn sequences based on Equation (\ref{new11}) & Notes \\
			\hline
			1 & (0000001111110111100011011001110100110000101110010101101010001001) & Equation (\ref{equ:nk1}) on index $t$  \\
			2 & (0000001100001010001111000100111010101101001011111101110011011001) &  \\
			3 & (0000001100110111111011010101100011110010111000010100111010001001) &  \\
			4 & (0000001111000110110011101111110100110000101110010101101010001001) &  \\
			5 & (0000001111011111100011011001110100110000101110010101101010001001) &  \\
			6 & (0000001100001010001111110111100010011101010110100101110011011001) &  \\
			7 & (0000001100110110101011000111111011110010111000010100111010001001) &  \\
			8 & (0000001111000110110011101001100001011111101110010101101010001001) &  \\
			9 & (0000001111000110111111011001110100110000101110010101101010001001) &  \\
			10 & (0000001100001010001111000100111011111101010110100101110011011001) &  \\
			11 & (0000001100110110101011000111101111110010111000010100111010001001) &   Equation (\ref{equ:nk1}) on index $s$\\
			\hline
			$k$ &  de Bruijn sequences based on Equation (\ref{equ:mix})  &   \\
			\hline
			$0$  &  $(0000001111110101011010010001001110110011011100101000010111100011)$  \\
			$1$  &  $(0000001100010010001111010101101000010100111011001101110010111111)$  \\
			$2$  &  $(0000001100110111001000100101111000111111010100001010110100111011)$  \\
			$3$  &  $(0000001101110010100001011111100011110101011010010001001110110011)$  \\
			$4$  &  $(0000001111110101011010000101001110110011011100101111000100100011)$  \\
			$5$  &  $(0000001100011110101000010101101001110110011011100100010010111111)$ \\
			$6$  &  $(0000001100110111001010000101111000111111010101101001000100111011)$  \\
			$7$  &  $(0000001101110010111111000100100011110101011010000101001110110011)$  \\
			$8$  &  $(0000001111110101000010101101001110110011011100100010010111100011)$  \\
			$9$  &  $(0000001100011110101011010010001001110110011011100101000010111111)$  \\
			$10$ &  $(0000001100110111001011110001001000111111010101101000010100111011)$  \\
			$11$ &  $(0000001101110010001001011111100011110101000010101101001110110011)$  \\
			\hline
		\end{tabular}
	\end{table*}
	\begin{example}\label{ex:nkorder}
		For $n=6$, Table \ref{table:FandG} lists $11$ generated distinct de Bruijn sequences from Proposition \ref{prop:nk}. Sequences from $k=1$ and $k=11$ are also those ruled by $\rho_s$ and $\rho_t$, respectively. \QEDB
	\end{example}
	
	\subsection{The Mixed Order on the Pure Summing Register}\label{subsec:mixed}
	
	This subsection imposes another new order on the cycles of the PSR of order $n$ to define novel successor rules. 
	
	We start with a new total order, called the {\it mixed order} and denoted by $\prec_{\rm mix}$, on the cycles of the PSR by \emph{combining the necklace order and the weight relation}. We say that $C_j \prec_{\rm mix} C_i$, if and only if they satisfy one of the following conditions.
	\begin{enumerate}[nolistsep]
		\item $\wt(C_j) > \wt(C_i)$.
		\item $\wt(C_j) = \wt(C_i)$ and, in their necklace order, $C_j \prec_{\rm nk} C_i$.
	\end{enumerate}
	
	It is rather typical in the construction of rooted spanning trees that adjacent cycles are chosen to follow lexicographically decreasing or increasing patterns. Adjacent cycles under our mixed order do not usually obey a lexicographic pattern. This sets our successor rules apart from those formulated in the spirit of the work by Sawada \etal in~\cite{Sawada17}.
	
	\begin{example}\label{ex:mix}
		In terms of weight, the $10$ cycles generated by the PSR of order $6$ have the following distribution. The weight of $(0)$ is clearly $0$. There are three cycles of weight $2$, five cycles of weight $4$ and one cycle of weight $6$. Given in increasing mixed order, the cycles are
		\begin{multline*}
			(0111111)\prec_{\rm mix} (0001111) \prec_{\rm mix} (0010111) \prec_{\rm mix} 
			(00011011)\prec_{\rm mix} (0011101) \prec_{\rm mix} \\
			(0101011) \prec_{\rm mix} (0000011) \prec_{\rm mix} (0000101) \prec_{\rm mix} (0001001)  \prec_{\rm mix} (0).
		\end{multline*}
		One confirms by inspection that the mixed order differs from the other orders that we have applied on these $10$ cycles. \QEDB
	\end{example}
	
	\begin{remark}
		There is another useful variant of the mixed order. We can say, alternatively, that
		$C_j \prec_{\rm mix} C_i$ if and only if they satisfy one of the following conditions.
		\begin{enumerate}[nolistsep]
			\item $\wt(C_j) < \wt(C_i)$,
			\item $\wt(C_j) = \wt(C_i)$ and, in the necklace order, $C_j \prec_{\rm nk} C_i$.
		\end{enumerate}
		This alternative also works in the next theorem, with the mechanism adjusted accordingly. \QEDB
	\end{remark}
	
	We give a sufficient condition based on the PSR of order $n$.
	\begin{theorem}\label{thm2}
		In the PSR of order $n$, let $\mathbf{c}:=c_0,c_1,\ldots,c_{n-1}$ be any given state and $\mathbf{v}:=0,c_1,\ldots,c_{n-1}$. The successor rule $\rho_{B}$ in (\ref{equ:PSR}) can be validly defined as follows.
		\begin{enumerate}[nolistsep]
			\item If there are two consecutive $0$s in $C:=\left(0,c_1,\ldots,c_{n-1},\sum_{i=1}^{n-1} c_i \right)$, the sum $\sum_{i=1}^{n-1} c_i=0$, and $\mathbf{v}$ can be uniquely determined, then $\rho_{B}(\mathbf{c}) := 1+ {\sum_{s=0}^{n-1} c_s}$.
			\item If there are no two consecutive $0$s in $C:=\left(0,c_1,\ldots,c_{n-1},\sum_{i=1}^{n-1} c_i \right)\neq(01^n)$ and exactly one of the followings holds, then  $\rho_{B}(\mathbf{c}) := 1+ {\sum_{s=0}^{n-1} c_s}$.
			\begin{enumerate}
				\item $\mathbf{v}$ is the only one LZ state.
				\item When there are two or more LZ states in $C$, $\mathbf{v}$ is uniquely identified and $0,c_1,\ldots,c_{n-1},1$ is not the necklace.
			\end{enumerate}  
			\item In all other cases, $\rho_{B} (\mathbf{c}) := {\sum_{s=0}^{n-1} c_s}$.
		\end{enumerate}
	\end{theorem}
	
	\begin{IEEEproof}
		The $(n+1)$-periodic cycle with the least mixed order is
		\begin{equation}\label{eq:C1}
			C_1 :=
			\begin{cases}
				(0,1,1,\ldots,1) \mbox{, if } n \mbox{ is even},\\
				(1,1,1,\ldots,1) \mbox{, if } n \mbox{ is odd}.\\
			\end{cases}
		\end{equation}
		By Theorem \ref{thm:order} and by the definition of $\rho_{B}$, it suffices to show that there exists a unique state in each $C\neq C_1$ whose successor, as determined by $\rho_{B}$, is contained in another cycle $C'$ with $C' \prec_{\rm mix} C$. It is clear that $C$ has at least one LZ state.
		
		Let $C$ be any cycle of the PSR that is not equal to $C_1$. Let there be two consecutive $0$s in $C$. Without loss of generality, let $C:=\left(0,c_1,\ldots,c_{n-1},\sum_{i=1}^{n-1} c_i \right)$ such that $\sum_{i=1}^{n-1}c_i = 0$, and $\mathbf{v}:=0,c_1,\ldots,c_{n-1}$ is the uniquely determined state. By the definition of $\rho_{B}$, the successor of $\mathbf{v}$ is $c_1,\ldots,c_{n-1},1$, which is in cycle $C':=(1,c_1,\ldots,c_{n-1},1)$. In this case, $\wt(C) < \wt(C')$ and, thus, $C' \prec_{\rm mix} C$, as desired. 
		
		If there are no two consecutive $0$s in $C$, then we can assume $C:=\left(0,c_1,\ldots,c_{n-1},1\right)$. If $C$ has only one LZ state, then the successor of $\mathbf{v}$ is $c_1,\ldots, c_{n-1},0$ by how $\rho_{B}$ is defined. If $C$ has multiple LZ states, then we choose $\mathbf{v}$ to be a uniquely determined LZ state such that $c_1,\ldots,c_{n-1},1$ is not the necklace. Again, by the definition of $\rho_{B}$, its successor is $c_1,\ldots, c_{n-1}, 0$. 
		
		In both cases, the successor is in another cycle $C':=\left(1,c_1,\ldots,c_{n-1},0\right)$ with $\wt(C) = \wt(C')$. By Lemma \ref{lem:15}, we have $C' \prec_{\rm nk} C$. Thus, $C' \prec_{\rm mix} C$, as required. By Theorem \ref{thm:order}, $\rho_{B}$ generates a de Bruijn sequence.
	\end{IEEEproof}
	
	Checking whether a cycle has two consecutive $0$s is very fast. Theorem \ref{thm2} implies that, if a cycle has two or more LZ states and the sum of the next $n-1$ bits in each of the states is $0$, then distinct LZ states lead to different successor rules. We supply several ways to uniquely identify qualified states.
	
	If a given cycle has two consecutive $0$s, then its necklace begins with two $0$s and all $(n+1)$-stage states that begin with two $0$s are obtainable by the shift operations on the necklace.
	We define a new operator, denoted by $L_{\rm dz}$. The subscript ${\rm dz}$ stands for {\it double zeros} to indicate that the operator is applicable on any state $\mathbf{v}:=0,0,c_1,\ldots,c_{n-1}$ that starts with two $0$s. The state $L_{\rm dz} \, \mathbf{v}$ is the first state with two leading $0$s obtained by repeated application of the left shift $L$ on $\mathbf{v}$. We now propose a new successor rule based on the mixed order.
	
	\begin{proposition}\label{prop:mix}
		Let $k$ be a nonnegative integer. For any state $\mathbf{c}:=c_0,c_1,\ldots,c_{n-1}$, let $C:=\left(0,c_1,\ldots,c_{n-1}, \sum_{i=1}^{n-1} c_i\right)$.
		Then
		\begin{equation}\label{equ:mix}
			\rho_{\rm dz}(\mathbf{c})=
			\begin{cases}
				1+\sum_{s=0}^{n-1}c_s \mbox{, if } \mathbf{c} \mbox{ satisfies Condition}\ X,\\
				\sum_{s=0}^{n-1}c_s \mbox{, otherwise,}
			\end{cases}
		\end{equation}
		generates a de Bruijn sequence. Condition $X$ consists of any of the following two requirements.
		\begin{quote}
			\begin{enumerate}
				\item The cycle $C$ has two consecutive $0$s, the sum $\sum_{i=1}^{n-1}c_i=0$, and $L_{\rm dz}^k \, \mathbf{v}$ is a necklace, where $\mathbf{v}:=0,0,c_1,\ldots,c_{n-1}$.
				\item The cycle $C \neq (0,1,1,\ldots,1)$ has no two consecutive $0$s and its necklace is $0,c_{j+1},\ldots,c_{n-1},1,0,c_1,\ldots,c_{j-1}$, with $1\leq j < n$ being the least positive integer for which $c_j=0$.
			\end{enumerate}	
		\end{quote}
	\end{proposition}
	
	\begin{IEEEproof}
		We prove this proposition by showing the successor rule $\rho_{\rm dz}$ satisfies the conditions in Theorem \ref{thm2} that a uniquely desired state is identified in each $C \neq C_1$, where $C_1$ is given in (\ref{eq:C1}).
		
		If $C$ has two consecutive $0$s and $\sum_{i=1}^{n-1}c_i=0$, then $C \neq C_1$ and $\mathbf{v}:=0,0,c_1,\ldots,c_{n-1}$ such that $L_{\rm dz}^k \, \mathbf{v}$ is a necklace is uniquely determined. consequently the state $0,c_1,\ldots,c_{n-1}$ is uniquely determined and its conjugate state is in $C':=(1,1,c_1,\ldots,c_{n-1})$, which has larger weight and, hence, $C' \prec_{\rm mix} C$.
		
		If $C\neq C_1$ has no two consecutive $0$s, then $C:=(0,c_1,\ldots,c_{n-1},1)$ must contain at least two $0$s. Hence, there is $j > 0$ such that $c_j=0$. Because $0,c_{j+1},\ldots,c_{n-1},1,0,c_1,\ldots,c_{j-1}$ is the necklace, $0,c_1,\ldots,c_{n-1}$ in $C$ is uniquely determined and, by Lemma \ref{lem:15}, its conjugate state must be in another cycle $C'$ with $C' \prec_{\rm nk} C$. Thus, $C' \prec_{\rm mix} C$.

	\end{IEEEproof}
	
	\begin{remark}
		There are many other ways to determine the unique state for the two cases in Statement $X$ of Proposition~\ref{prop:mix}. Here is an example. If $C$ contains two consecutive $0$s, then we can use the method proposed in Subsection \ref{subsec:neck} to determine the unique state and, hence, to provide new successor rules.
	\end{remark}
	
	\begin{proposition}
		The successor rules $\rho_{\rm dz}$ in (\ref{equ:mix}), identified with different valid choices for $k$, generate
		\begin{equation}\label{eq:lbounds}
			\lcm(1,2,\ldots,n-2) 
			\geq  (n-2)  \binom{n-3}{\left\lfloor \frac{n-3}{2} \right\rfloor}  \geq 2^{n-3}
		\end{equation}
		distinct de Bruijn sequences in total.
	\end{proposition}
	\begin{IEEEproof}
		It suffices to count the number of two consecutive $0$s in the cycles that contain consecutive $0$s. For each $2 \leq i < n$, there exists at least one cycle that contains $i$ consecutive $0$s. For example, when $i=n-1$, the cycle $(0,\ldots,0,1,1)$ has a run of $0$s of length $n-1$. For each $1 \leq j \leq n-2$, there is at least one cycle that has $j$ distinct states, each of which can be declared to be the uniquely determined state. The same reasoning we used in proving Proposition~\ref{prop2} tells us that each $k$ in $0 \leq k < \lcm(1,2,\ldots,n-2)$ yields a de Bruijn sequence. Different values for $k$ produce distinct sequences based on the corresponding successor rules. The lower bounds in (\ref{eq:lbounds}) comes from \cite{Farhi}.
	\end{IEEEproof}
	
	\begin{example}\label{ex:mix2}
		Table~\ref{table:FandG} contains $12$ distinct de Bruijn sequences produced by using  Proposition~\ref{prop:mix} with $n=6$.\QEDB
	\end{example}
	
	
	\section{Conclusions}\label{sec:conclu}
	
	In this paper, we have proposed a general design criteria for feasible successor rules. They perform the cycle joining method to output binary de Bruijn sequences. The focus of our demonstration is on their efficacy and efficiency when applied to two classes of simple FSRs. These are the pure cycling register (PCR) and the pure summing register (PSR) of any order $n \geq 3$. 
	
	Our approach is versatile. It goes beyond the often explored route of relying on the lexicographic ordering of the cycles. We have shown that many transitive relations can also be used to put the cycles in some order. We have enumerated the respective output sizes of various specific successor rules that can be validly defined based on the general criteria. A straightforward complexity analysis has confirmed that generating the next bit in each resulting sequence is efficient.
	
	We assert that the criteria we propose here can be applied to \emph{all nonsingular FSRs}. If a chosen FSR has cycles with small least periods, then the complexity to produce the next bit can be kept low. Interested readers are invited to come up with feasible successor rules for their favourite FSRs. We intend to do the same and to further look into, among others, the cryptographic properties of the binary de Bruijn sequences produced by more carefully designed successor rules.
	
	\section*{Acknowledgements}
	We thank Adamas Aqsa Fahreza and Yunlong Zhu for various {\tt python} and {\tt C} routines that greatly helped our investigations. The collaboration leading to this paper commenced when Zuling Chang was hosted by Khoa Nguyen at the School of Physical and Mathematical Sciences, Nanyang Technological University, in July 2019.
	
	

\end{document}